\documentclass[pra,twocolumn,preprintnumbers,amsmath,amssymb,floatfix,superscriptaddress,showpacs]{revtex4}
\usepackage[dvips]{graphicx}
\usepackage{graphics}
\usepackage{psfrag}
\usepackage{hyperref}

\begin{document}

\hypersetup{pdftitle={Hydrodynamic collective modes for cold trapped gases}}
\title{Hydrodynamic collective modes for cold trapped gases}
\date{\today}
\author{Igor Boettcher}
\affiliation{Institut f\"{u}r Theoretische Physik, Universit\"{a}t Heidelberg,\\Philosophenweg 16, D-69120 Heidelberg, Germany}
\author{Stefan Floerchinger}
\affiliation{Institut f\"{u}r Theoretische Physik, Universit\"{a}t Heidelberg,\\Philosophenweg 16, D-69120 Heidelberg, Germany}
\affiliation{Physics Department, Theory Unit, CERN, CH-1211 Gen\`{e}ve 23, Switzerland}
\author{Christof Wetterich}
\affiliation{Institut f\"{u}r Theoretische Physik, Universit\"{a}t Heidelberg,\\Philosophenweg 16, D-69120 Heidelberg, Germany}

\begin{abstract}
We suggest that collective oscillation frequencies of cold trapped gases can be used to test predictions from quantum many-body physics. Our motivation lies both in rigid experimental tests of theoretical calculations and a possible improvement of measurements of particle number, chemical potential or temperature. We calculate the effects of interaction, dimensionality and thermal fluctuations on the collective modes of a dilute Bose gas in the hydrodynamic limit. The underlying equation of state is provided by non-perturbative Functional Renormalization Group or by Lee--Yang theory. The spectrum of oscillation frequencies could be measured by response techniques. Our findings are generalized to bosonic or fermionic quantum gases with an arbitrary equation of state in the two-fluid hydrodynamic regime. For any given equation of state $P(\mu,T)$ and normal fluid density $n_n(\mu,T)$ the collective oscillation frequencies in a $d$-dimensional isotropic potential are found to be the eigenvalues of an ordinary differential operator. We suggest a method of numerical solution and discuss the zero-temperature limit. Exact results are provided for harmonic traps and certain special forms of the equation of state. We also present a phenomenological treatment of dissipation effects and discuss the possibility to excite the different eigenmodes individually.
\end{abstract}


\maketitle


\section{Introduction}
Ultracold quantum gases provide a powerful experimental tool to investigate many-body phenomena in quantum systems \cite{BDZ,PiSt1,PeSm1}. Their distinguished role arises from mainly two facts. First they constitute clean realizations of microscopic Hamiltonians. For example, whereas a continuum gas of fermions with contact interaction or the (Bose--)\ Hubbard-model are only approximations to solid state systems, they provide realistic descriptions for cold trapped gases. Therefore the latter allow for a direct verification of theoretical predictions concerning these simple microscopic models. The second important feature of ultracold gases is the possibility to tune the system parameters by application of external fields and to enter the regime of strong interactions with the help of Feshbach resonances. Again, this is in clear contrast to e.g. solids, where the parameters are fixed by the sample.

An important and promising direction in cold gases research is the study of strongly correlated systems. The latter arise on all scales of energy, ranging from high-temperature superconductors to neutron stars, heavy-ion collisions and the phase structure of quantum chromodynamics. These systems require sophisticated theoretical methods, because mean-field theory and perturbative techniques break down due to the strong correlations. However, since there is no small parameter, the accuracy of non-perturbative approaches is far from being well understood. The systematic comparison with experiments could therefore shed light on the reliability of first-principle methods like lattice simulations, the renormalization group,  2PI approaches, extensions of density functional theory, conformal field theory etc.

From our above consideration it should be clear that ultracold quantum gases provide an ideal basis for such a systematic comparison of theory and experiment. We are thus lead to an important question: What observables for cold gases reveal advancements and shortcomings of our theoretical treatment of strongly correlated systems? The measurement of these quantities will then set stringent bounds on theoretical descriptions and rule out insufficient methods. 

In recent experiments, the equation of state has proven to be such an observable \cite{Nav2,Nav1}. In this paper we investigate to what extend collective oscillation frequencies of trapped gases, which can be derived from the equation of state, contain similar or additional information. For a trapped gas collective modes describe small deviations from static equilibrium. They typically show an oscillatory behavior with characteristic frequencies. The number density of atoms in a trap obeys for a particular oscillation mode
\begin{equation}
 \label{int-1} n(\vec{x},t) = n_0(\vec{x}) + \delta n(\vec{x}) \mbox{cos}(\omega t), 
\end{equation}
where $n_0(\vec{x})$ denotes the equilibrium density profile and we assume small amplitudes, $\delta n \ll n_0$. We will show how $\omega$ can be calculated for a $d$-dimensional trap in the ideal two-fluid hydrodynamic regime for an arbitrary equation of state.

For a gas in equilibrium the equation of state consists in the pressure $P$ as a function of two independent thermodynamic variables, e.g. chemical  potential $\mu$ and temperature $T$. In situations where local density approximation (LDA) is valid, the equation of state $P(\mu,T)$ can be extracted from density images of the trapped cloud \cite{Ho2}. We will specify the applicability of LDA below. Here we mention only that the system has to be considered in equilibrium at each point of the trap separately. This might seem a strong statement at first sight, but, for sufficiently high densities and strong interactions, experiments have shown LDA to be valid at least approximately. For example, the coefficient of the Lee--Huang--Yang-correction to the equation of state of a bosonic gas could be measured to good accuracy \cite{Nav1}. Since the equation of state is obtained from a calculation of the partition function, particular numbers like critical temperature and exponents, or even the whole equation of state $P(\mu,T)$ over a certain range of $\mu$ and $T$, can be used to verify predictions from many-body theory. If a theoretical method fails in the context of cold atoms, it might also be questionable in other situations.

For sufficiently low temperatures ultracold quantum gases can show superfluidity. Then the equation of state consists in two functions, the pressure $P(\mu,T)$ and the normal fluid density $n_n(\mu,T)$. Again these quantities can be computed from first principles.

The response of a collision-dominated system in thermodynamic equilibrium to a slight perturbation away from equilibrium is completely determined by conservation of mass, energy and momentum. To first order this response can be computed from ideal hydrodynamics and thus only requires knowledge of the equation of state -- an equilibrium quantity. Since thermodynamics and thus LDA is the static limit of ideal hydrodynamics, the applicability of the latter might also be given for sufficiently high densities and strong interactions. The question whether the hydrodynamic limit can be reached in experiments is under debate. However, this regime is very promising from a theoretical perspective because details about the experimental setting become less important. Therefore, already a few quantities like the equation of state or transport coefficients can be used to test theory. One of the purposes of this paper is to provide an idea how measurements in this regime could look like in future experiments. If it should turn out to be impossible to reach the ideal hydrodynamic regime of cold gases, our considerations are still valuable as limiting cases of more elaborate treatments.

In the presence of a non-vanishing superfluid density the ideal hydrodynamic equations have to be modified because the order parameter enters as an additional macroscopic degree of freedom. The proper two-fluid hydrodynamic equations have been derived by Landau. We propose that collective modes of trapped gases in this regime are an interesting observable in the above sense, i.e. they can be used to test theoretical methods which calculate the equation of state from a given Hamiltonian. The measurement of oscillation frequencies can supplement direct determinations of the equation of state from density images. In fact, we will show in this paper how the spectrum of collective modes can be computed from the knowledge of the functions $P(\mu,T)$ and $n_n(\mu,T)$ by solving an eigenvalue problem. Since thermodynamic derivatives up to second order enter the calculation, insufficient calculations of the equation of state will not yield consistent results for several modes and over wide ranges of external parameters like temperature and particle number. Of particular interest is the behavior of the equation of state at the second order superfluid phase transition.

Any observable which is well understood as a function of external parameters like temperature, particle number or chemical potential can, of course, in turn be used to measure these parameters. Since the hydrodynamic regime is insensitive to microscopic details, collective modes can yield information about macroscopic properties of the system. Several conditions have to be fulfilled for such an investigation to be reasonable in the first place: The hydrodynamic regime has to be reached in experiment, the equation of state has to be known to sufficient accuracy and the mapping between the equation of state and collective oscillation frequencies has to be understood. Contributing to the latter point is the subject of this paper.

Collective oscillation frequencies have been a promising observable since the first experiment on trapped cold gases. Therefore, a great amount of literature already exists on measurement \cite{Jin1, SK1,Che,Ess,Kin1,Alt1,Kott,Poll} and calculation \cite{Edw,Str1,Per,Str2,Str3,Hut,Ho1,DoEd1,BrCl1,Hu,Gio,Jac,KiZu,Ast1,Tay1} of these modes for weakly interacting Bose gases, the BEC-BCS crossover and other systems at both zero and nonzero temperature. While earlier work \cite{Fed, ZaNi, BiSt1, JaZa} has partly taken into account additional microscopic properties, we concentrate here on the hydrodynamic approximation.

The main results of this paper are the following. We derive the eigenvalue problem which determines the hydrodynamic modes for a given equation of state $P(\mu,T)$ and $n_n(\mu,T)$ in a $d$-dimensional trap. A numerical implementation based on discretization is put forward. To be concrete we consider an isotropic harmonic trap in our calculations. However, the extension to arbitrary isotropic traps is straightforward, an implementation of anisotropic external potentials requires more numerical effort but is in principle possible. 

For a zero temperature Bose gas, where hydrodynamics is expected to be valid because the system is completely superfluid, we calculate the correction of the breathing mode beyond mean-field in three and two dimensions. The equation of state $P(\mu)$ in this case is obtained from non-perturbative Functional Renormalization Group. We find the expected shift of the breathing mode as a function of the gas parameter in three dimensions and provide a similar result in two dimensions, where the gas parameter has to be defined differently. We emphasize however that our method is not restricted to the lowest breathing mode, but rather allows to determine the full spectrum of oscillation frequencies including higher dipole, quadrupole, etc. modes. 

As a generic example at nonzero temperature we consider a dilute Bose gas described by Lee--Yang-theory. We determine the frequencies over a wide range of temperature, varying from low temperatures to the normal regime above the superfluid transition temperature. Although a dilute Bose gas is not necessarily expected to fulfill the conditions of hydrodynamics, our method can formally be applied and allows to extract generic features that could arise in spectra of oscillation frequencies. 

We discuss the zero temperature limit in the case of a Bose gas and find a new set of zero temperature frequencies due to oscillations of the normal atoms which cannot be resolved by purely superfluid hydrodynamics of the condensate. We explain why it is important to measure several frequencies and how this can be achieved with response techniques.

Due to the narrow constraints of thermodynamics the equation of state and thus the collective modes of other systems are expected to show a similar behavior. Particularly interesting is a strongly interacting Fermi gas, where, however, the equation of state is not yet known sufficiently, especially below the critical temperature. Further experimental and theoretical progress in this direction will allow for the calculation of the collective modes of a unitary Fermi gas as a function of temperature. 

Our paper is organized as follows. In Sec. \ref{sec1} we present our results on the dilute Bose gas. We are interested in contact to experiment and focus on easy readability, giving only the relevant equations without derivation. The reader will find references to the subsequent sections, where derivations of the formulas are presented in a general framework. In Sec. \ref{coll} we derive the eigenvalue problem for an arbitrary equation of state at zero and non-vanishing temperature. This will yield the formulas already used in Sec. \ref{sec1}. The numerical implementation to obtain the collective frequencies, which is used throughout this paper, is given in Sec. \ref{num}. We then come to our conclusion in Sec. \ref{concl}. In App. \ref{unit} our chosen system of units is explained and App. \ref{appLY} summarizes the theory of Lee and Yang for the dilute Bose gas. In App. \ref{appC} spherical harmonics in $d \leq 3$ dimensions are introduced. Exact results for the zero temperature Bose gas are derived in App. \ref{exact}. Details of our phenomenological discussion of response techniques are given in App. \ref{appE}.

\section{Oscillations of a Bose gas}
\label{sec1}
In this section we demonstrate our ideas for the example of a Bose gas. Due to the simplicity of the latter we expect our findings to be generic for a broad class of thermodynamic quantum systems. We keep our presentation very brief at this point and refer the reader to the next section where the general theory of collective oscillations is provided in detail.

At low temperatures and low densities interactions in a homogeneous Bose gas can be described completely in terms of contact interactions. In this regime the precise form of the interaction potential is irrelevant, implying microscopic universality. In three dimensions the associated coupling constant $g_{3D}$ has dimension of length and is related to the s-wave scattering length $a$ through $g_{3D}=4 \pi \hbar^2 a/m$ for bosons with mass $m$. For composite bosons all formulas remain valid with small modifications. For example, if we aim at describing the weakly interacting BEC-side of the BEC-BCS crossover for two component fermions, $a$ has to be replaced by the dimer-dimer scattering length, which is proportional to the fermion scattering length \cite{Pet1}, and $m$ becomes the dimer mass, which is two times the fermion mass. This correspondence for the equation of state has been nicely demonstrated in Ref. \cite{Nav1}. In two dimensions the coupling constant becomes dimensionless. The effects arising in this interesting situation will be investigated later in this section.

\subsection{Three-dimensional dilute Bose gas at zero temperature}

For a three-dimensional Bose gas the condition $na^3 \ll 1$, where $n$ is the density, corresponds to a dilute and weakly interacting system. In this case, the mean-field result for the energy density at zero temperature is given by \cite{Bog1,PiSt1,PeSm1} ($\hbar=k_B=m=1$, see App. \ref{unit} for units)
\begin{equation}
 \label{2-0} \varepsilon(n,a) =\frac{g_{3D}}{2} n^2 = 2 \pi a n^2.
\end{equation}
Taking a derivative with respect to $n$, we get the chemical potential $\mu=g_{3D}n$ which enters the Gibbs-Duhem relation $\mbox{d}P=n \mbox{d}\mu$ for the pressure. The equation of state written in the grand canonical variables then becomes
\begin{equation}
 \label{2-1} P(\mu,a) = \frac{\mu^2}{8 \pi a},
\end{equation}
which is of polytropic type $P\propto \mu^{\alpha+1}$ with $\alpha=1$. At zero temperature the system is completely superfluid and will be described by superfluid hydrodynamics.

The oscillation frequencies of this system in a spherical parabolic trap $V_{ext} = \frac{m}{2} \omega_0^2 r^2$ are found by solving the eigenvalue problem
\begin{equation}
 \label{2-2} A g(z) = \left(\frac{\omega}{\omega_0}\right)^2 g(z)
\end{equation}
for the differential operator
\begin{equation}
 \label{2-2a} A = -\frac{P^\mu(z)}{P^{\mu\mu}(z)} \left(4 z \frac{\partial^2}{\partial z^2} + 2(2 l +d)\frac{\partial}{\partial z}\right) + \left(2 z \frac{\partial}{\partial z} +l \right)
\end{equation}
acting on a function $g(z)$. Here $d=3$, see Sec. \ref{coll}. The operator $A$ depends on the equation of state through $P^\mu(\mu)$ and $P^{\mu\mu}(\mu)$, where a superscript denotes differentiation with respect to $\mu$. The former of these quantities corresponds to the density $P^\mu=n$ while their ratio $P^\mu/P^{\mu\mu} = \partial n/\partial P=c^2$ is related to the velocity of sound $c$. Since only this ratio appears in the operator $A$, the prefactor of Eq. (\ref{2-1}) is of no importance and the frequencies will not depend on the interaction strength on the mean field level. Indeed, Stringari \cite{Str1} found the analytic expression $\omega_{nl} = (2n^2 + 2nl +3n +l)^{1/2}\omega_0$, where $n$ and $l$ are integers. Obviously, no thermodynamic conclusions can be drawn from this formula and it may at best help to determine the trapping frequency when interactions and thermal effects are known to be very small. Stringari's formula is a special case of
\begin{equation}
\label{2-2b} \omega_{\alpha,n,l} = \left( \frac{2n}{\alpha} \left(\alpha + n +l +d/2 - 1\right) + l\right)^{1/2} \omega_0,
\end{equation}
which holds for $d$-dimensional spherically symmetric, harmonic traps and arbitrary polytropic index $\alpha$. As we will show in App. \ref{exact}, this formula can be obtained by applying a simple polynomial ansatz $g(z)=\sum_{k=0}^n a_k \bar{z}^k$ with $\bar{z}=z/R^2$ to Eq. (\ref{2-2}) for $P\propto \mu^{\alpha+1}$. It is already known in the literature \cite{Fuc1,Hei1}. 

Eq. (\ref{2-2b}) reveals that the possible oscillation frequencies of a trapped system form a discrete set. We will see in the following that this property is true for an arbitrary equation of state $P(\mu,T)$. This feature does not arise as a result of imposed boundary conditions but is related to the domain of definition of the operator $A$ in a different way. We will discuss this issue in Sec. \ref{num} where we solve the eigenvalue problem (\ref{2-2}) on a finite grid by discretization. Here, we only remark that the indices $n$ and $l$ of Eq. (\ref{2-2b}) have the same meaning as in the probably more familiar case of a quantum mechanical particle in a spherical symmetric potential. Similar to the hydrogen atom, $n$ is related to the number of nodes of the collective mode, where $g(z)$ plays a role analogous to the wave function. Rotation symmetry implies a conserved angular momentum $l$ which enters the operator $A$ as a parameter. For $l=0$ the collective motion is isotropic, while for $l=1$ it is dipolar, and so on. Details can be found in Sec. \ref{coll} where we derive Eq. (\ref{2-2a}). We expect formula (\ref{2-2b}) to be applicable only for sufficiently small values of $n$ and $l$, because many nodes or a complicated angular structure may enter in conflict with the assumptions of hydrodynamics. 

Apparently, for each $l>0$ there is a mode with $n=0$ and frequency $\sqrt{l}\omega_0$ which is independent of the equation of state. In particular, the lowest dipole mode is exactly at the trapping frequency. Up to now many measurements focused on the breathing and lowest quadrupole mode, $n=1, l=0$ and $n=1,l=2$, respectively. We emphasize that the measurement of two, three or more frequencies could achieve high precision when combined with theory: For a given equation of state an arbitrary number of eigenfrequencies can be obtained by our numerical procedure. Comparison to measurements in well-understood parameter regions tests the ability of a given theoretical method to compute the equation of state accurately. On the one hand, it will at least give us an estimate on the theoretical errors. On the other hand, if the agreement is very good, we can take advantage of a non-trivial dependence of the frequencies on thermodynamic quantities like temperature, density, chemical potential or equation of state parameters like $a$. For example, if we predict the lowest lying three monopole modes to have a specific temperature dependence, we may conclude from measuring these three modes in what temperature region we are. 

Motivated by this we are looking for non-trivial relations between collective modes and thermodynamic quantities. We have already seen that a system purely described by mean field theory does not show such a behavior. We expect the mean field picture to be valid for very small gas parameter $na^3\ll1$. However, as this parameter increases, higher order interaction effects become relevant in the equation of state. The leading order correction to the ground state energy density (\ref{2-0}) at zero temperature has been calculated by Lee, Huang and Yang and is found to be \cite{LHY1}
\begin{equation}
 \label{2-3} \varepsilon_{LHY}(n,a) = 2 \pi a n^2\left(1 + \frac{128}{15 \sqrt{\pi}} (na^3)^{1/2}\right).
\end{equation}
The corresponding pressure as a function of chemical potential to the same order in perturbation theory is given by
\begin{equation}
\label{2-4} P(\mu,a) = \frac{\mu^2}{8 \pi a} - \frac{8}{15 \pi^2} \mu^{5/2}.
\end{equation}
Apparently, the interaction strength can no longer be eliminated by a rescaling of $P(\mu)$ such that the ratio $P^\mu/P^{\mu\mu}$ entering the operator $A$ will always depend on $a$. Thus, the frequencies will depend on the coupling constant. While for a homogeneous system the thermodynamic observables are constant in space, for a confined gas one usually refers to their values at the center of the trap. Denoting the density in the center of the external potential by $n(0)$, $a^3 n(0)$ provides a small parameter and one expects a shift of the breathing mode ($n=1,l=0$, i.e. $\omega_B=\sqrt{5}\omega_0$) \cite{PiSt2,Bra1}
\begin{equation}
 \label{2-5} \frac{\delta \omega_{B}}{\omega_{B}} = \frac{63 \sqrt{\pi}}{128} \left[a^3 n(0)\right]^{1/2}
\end{equation}
as compared to the Stringari formula above. We conclude that we may use the shift for small values of $n(0)a^3$ in order to determine either $n(0)$ or $a$ very precisely if the corresponding second quantity is known from another method. 

\begin{figure}[tb!]
\includegraphics[scale=1,keepaspectratio=true]{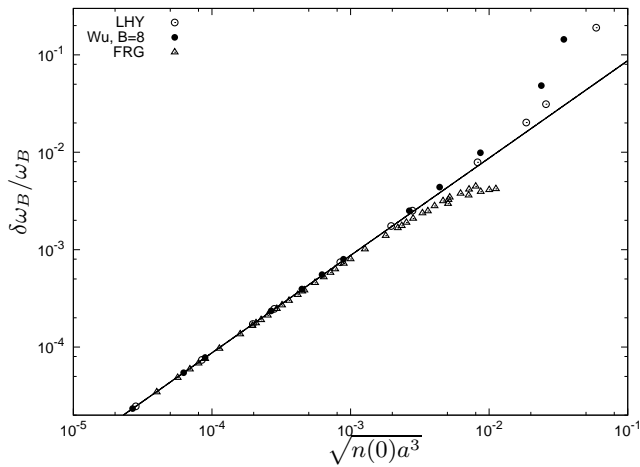}
\caption{Shift of the mean-field breathing mode $\omega_B=\sqrt{5}\omega_0$ for the zero temperature Bose gas in three dimensions. We show predictions for different methods: Lee--Huang--Yang-formula (\ref{2-4}), Wu-correction (\ref{2-6}) with $B=8$ and equation of state from Functional Renormalization Group \cite{Flo1} (FRG). The solid line corresponds to Eq. (\ref{2-5}), $a$ is the scattering length and $n(0)$ the density in the center of the cloud.}
\label{fig1}
\end{figure}

From this example we can deduce a method to make high precision measurements using collective modes. If we assume the shifts of the frequencies $\omega_{n,l}$ to be continuous in $n(0)a^3$ or $\mu(0) a^2$, where $\mu(0)$ is the chemical potential in the center of the trap, then for small gas parameters, $n(0)a^3\ll1$ or $\mu(0) a^2 \ll1$, the shift will be proportional to some power of the gas parameter. In the case of Eq. (\ref{2-5}) we found this power to be $1/2$. However, this will depend on the system under consideration. Nevertheless, driving the system through a certain range of the gas parameter, e.g. by the use of a Feshbach Resonance, and then collecting the result in a double-logarithmic plot one will find the results to lie on a straight line for small gas parameter. This simple scaling behavior can be used for very precise measurements of $n(0)a^3$ after a proper calibration. The accuracy of this method is directly related to the accuracy in measuring the frequency shifts. In Fig. \ref{fig1} we show the shifts beyond mean field due to the LHY-correction which we obtained by our numerical implementation described in Sec. \ref{num}. We reproduce the prediction to a high precision in the regime of small gas parameter, where it is valid. Although this plot only shows the breathing mode we emphasize that it is possible to calculate the shifts of the whole frequency spectrum which are found to show a similar scaling behavior with $\sqrt{n(0)a^3}$. Fig. \ref{fig1} shows that an experimental accuracy for frequency shifts better than $10^{-2}$ would be useful for exploring the regime of small gas parameters.

The calculation of higher order corrections to the LHY equation of state (\ref{2-3}) is a difficult task. The next-to-next-to-leading order has been derived by Wu \cite{Wu1,Hua1} for hard-sphere bosons. The energy density receives two additional terms
\begin{equation}
 \label{2-6}  \varepsilon_{LHY} + 2 \pi a n^2\left(8\left(\frac{4\pi}{3}-\sqrt{3}\right)na^3\right)(\log{(na^3)} + B).
\end{equation}
Neglecting the logarithmic term, this leads to a correction which is proportional to $\mu^3 a$ in Eq. (\ref{2-4}). In Fig. \ref{fig1} we have set $B=8$, see e.g. Ref. \cite{Nav1} for an overview over the expected values of $B$. Obviously, the general behavior of the frequency shifts is not influenced. 

For large values of the gas parameter perturbation theory is no longer applicable and more sophisticated methods are necessary. The Functional Renormalization Group (FRG) for the effective average action \cite{Wet1} is a non-perturbative quantum field theoretical method and is in particular able to describe the full thermodynamics of systems which are realized in cold atom experiments. It does not rely on the expansion in a small parameter and therefore the most striking effects are expected in strongly interacting systems. For more information on the FRG see Refs. \cite{FRG}. It is still under debate whether the regime $a \rightarrow \infty$ can be reached for a Bose gas at very low temperatures. One expects the condensate to get unstable then because of increasing importance of three-body losses. 

In Fig. \ref{fig1} we show the frequency shift obtained from the equation of state calculated with FRG at zero temperature \cite{Flo1}. We see that higher values of the gas parameter $n(0) a^3$ can no longer be identified with a unique $\delta \omega_B/\omega_B$. This is a fully non-perturbative effect. We recall that one of the assumptions in the beginning was that the interaction can be approximated to be point-like. However, this is only valid if the scattering length $a$ is much larger than the microscopic distance $\Lambda^{-1}$ where one can resolve the details of the interaction potential. For cold bosonic atoms this length is given by the typical range of the Van der Waals potential. An effective ultraviolet cutoff $\Lambda$ is  necessary in any field theoretic treatment of the system where the contact interaction appearing in the Lagrangian of the non-relativistic system is not renormalizable in three dimensions. Therefore, the introduction of this scale is no relict of approximations but has a clear physical meaning. In perturbation theory one assumes $\Lambda$ to be infinity and thus the equation of state of the homogeneous system only involves one dimensionless parameter, the gas parameter $na^3$. A proper treatment has to account for the appearance of an additional length scale $\Lambda^{-1}$ such that two possible combinations, $na^3$ and $a\Lambda$, describe the system. As a result the shift of the breathing mode should rather be plotted as a two-dimensional surface depending on these two parameters. For small gas parameter the additional parameter $a \Lambda$ does not play a role -- if we take equations of state $P(\mu,a)$ for different values of $a$ and then vary $n(0)$, all obtained shifts will lie on a line in the double-logarithmic plot. This is the expected scaling behavior. For higher densities it makes a difference whether we calculate the frequencies from $P(\mu,a, \Lambda_1)$ or $P(\mu,a,\Lambda_2)$ with $\Lambda_1 \neq \Lambda_2$, as reflected by the spread of the points in Fig \ref{fig1}.

\subsection{Lower-dimensional systems}

With regard to formula (\ref{2-2b}) we may wonder whether the cases $d=2$ and $d=1$ refer to highly anisotropic traps. These are of great relevance for experiments where one is often not dealing with a spherical symmetric potential. For example, the early experiments in the JILA group were performed in a disk-shaped confinement \cite{Jin1} while the MIT group used a cigar-shaped trap \cite{SK1}.

A really lower-dimensional system is obtained in a trap where the quantum gas is in its ground state in one or two directions. This will be the case for very tight confinement. When calculating the equation of state one can then neglect quantum and thermal fluctuations in these directions. The mean field Bose gas in this scenario is described by Eq. (\ref{2-2b}) when inserting $d=2$ or $d=1$ and $\alpha=1$. In these cases the system is isotropic in a lower-dimensional geometry. 

A different situation arises if the system is not in its ground state in the directions of tight confinement and thus still three-dimensional or in a crossover between three and lower dimensions. Formula (\ref{2-2b}) cannot be applied in this case because the assumption of spherical symmetry is not justified. The solution of the hydrodynamic equations for these anisotropic cases is more involved.

We now consider the two-dimensional Bose gas. Its mean field equation of state is $\mu = g_{2D} n$ and the frequency of the breathing mode is found from Eq. (\ref{2-2b}) to be $\omega_B=2 \omega_0$. The equation of state beyond mean field can be obtained by a Functional Renormalization Group approach \cite{Flo1}. The coupling constant $g_{2D}$, which satisfies $\mu = g_{2D}n$ for small $g_{2D}$, is dimensionless in two dimensions and shows a logarithmic running with the physical scale on which the experiment is performed. It will vanish for an infinitely large system. However, realistic experiments are always performed in traps so that in a harmonic potential the oscillator length $\ell_0=\sqrt{\hbar/m\omega_{xy}}$ provides the largest possible length scale of the physics under considerations. Therefore, when calculating the equation of state one only has to include quantum fluctuations with momenta bigger than $\ell_0^{-1}$, which acts as an infrared cutoff. This is also the reason why Bose--Einstein condensation is observed experimentally in two dimensions for small temperatures $0 < T < T_{c}$. For an infinitely extended system the long-range order would be destroyed by fluctuations for all non-vanishing temperatures as required by the Mermin-Wagner theorem \cite{Mer1}.

\begin{figure}[tb!]
\includegraphics[scale=1,keepaspectratio=true]{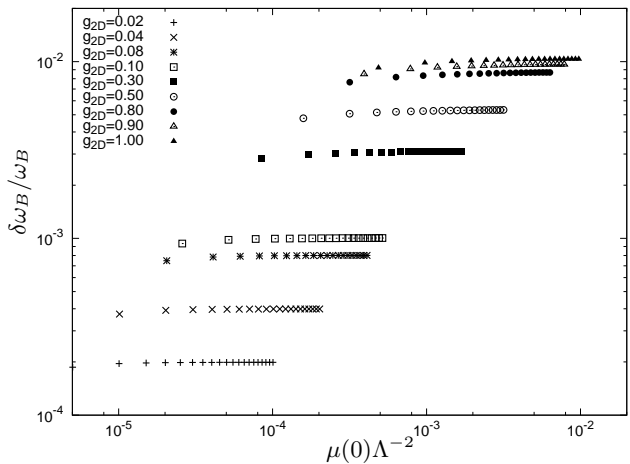}
\caption{Shift of the breathing mode relative to $\omega_B=2 \omega_0$ for a two-dimensional dilute Bose gas for different values of the coupling constant $g_{2D}$. We show the dependence on $\mu(0) \Lambda^{-2}$, with chemical potential $\mu(0)$ in the center of the trap and effective UV cutoff $\Lambda$. The equation of state is provided by Functional Renormalization Group.}
\label{fig2}
\end{figure}

\begin{figure}[tb!]
\includegraphics[scale=1,keepaspectratio=true]{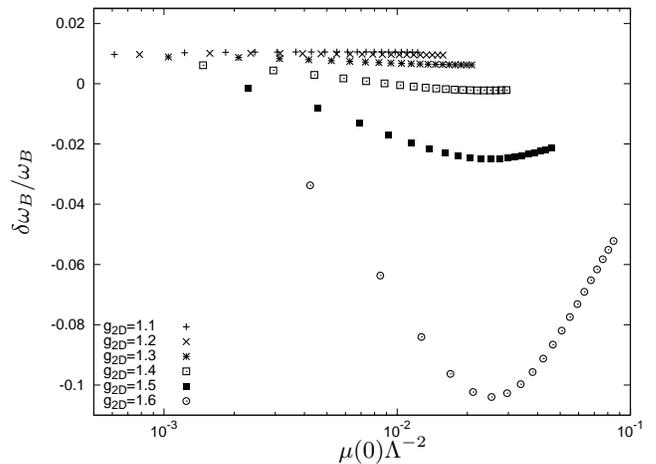}
\caption{The same setting as in Fig. \ref{fig2} but for higher values of the coupling $g_{2D}$. We find rather large negative shifts, which have a minimum.}
\label{fig3}
\end{figure}

In Figs. \ref{fig2} and \ref{fig3} we plot the shift of the breathing mode corresponding to $\omega_B=2 \omega_0$ for the two-dimensional Bose gas at zero temperature with equation of state $P(\mu,g_{2D})$ from Functional Renormalization Group calculations \cite{Flo1}. Since the coupling constant is dimensionless there is no gas parameter for such a system. As we mentioned already in the three-dimensional case one always has a physical ultraviolet cutoff $\Lambda$ when dealing with a contact interacting in $d\geq2$ dimensions. Since $\Lambda$ has dimension of inverse length, the dimensionless variable involving the chemical potential (or similar for the density) is $\mu(0) \Lambda^{-2}$, where $\mu(0)$ is the chemical potential of the gas in the center of the trap. A good choice for $\Lambda^{-1}$ is the oscillator length of tight trapping. We observe from Fig. \ref{fig2} that the frequencies for small interactions depend only weakly on $\mu(0) \Lambda^{-2}$. For larger coupling $g_{2D}$ we find deviations from this behavior in Fig. \ref{fig3}. We emphasize that the dependence on the microscopic cutoff for two-dimensional bosonic systems is not a shortcoming of approximations, but a physical effect. It is due to the running of couplings. While the relevant coupling is dimensionless, the equation of state retains memory of scales in form of a dependence on the physical effective cutoff. This effect becomes substantial for large $g_{2D}$. Therefore, the strong dependence of the frequency shift on the dimensionless combination $\mu(0) \Lambda^{-2}$ for large interaction strength $g_{2D}\gtrsim 1$ in Fig. \ref{fig3} signals the breakdown of universality. More microscopic properties beyond the single interaction parameter $g_{2D}$ are needed to describe a concrete experimental realization in this regime.

We arrive at two interesting experimental scenarios to be investigated. On the one hand, by measuring the collective modes one can distinguish whether one is working with a system which is still three-dimensional (disk, cigar) or a truly lower dimensional system. On the other hand, if the latter regime is reached experimentally it is of course tempting to verify the predictions of Fig. \ref{fig3}. The rather large negative frequency shifts can both test theoretical predictions like the occurrence of a minimum, and determine $\mu(0)$ or the corresponding density $n(0)$.

\subsection{Three-dimensional dilute Bose gas for non-vanishing temperature}

We now extend our considerations to non-vanishing temperature and allow for fluctuations of both density and temperature. If we stay below the critical temperature of Bose--Einstein condensation the system possesses both non-vanishing superfluid and normal fluid density and its hydrodynamic behavior has to be described by two-fluid hydrodynamics \cite{Lan1, Kha1, Put1}.  

As explained in Sec. \ref{coll}, we again find an eigenvalue problem which has to be solved in order to get the collective frequencies of the system. It has the general form
\begin{equation}
 \label{2-9} \left(\begin{array}{cc}
	A & B \\ C & D
      \end{array}
\right)
\left(\begin{array}{c}
       g(z) \\ h(z)
      \end{array}
\right)
= \left(\frac{\omega}{\omega_0}\right)^2
\left(\begin{array}{c}
       g(z) \\ h(z)
      \end{array}
\right),
\end{equation}
where the differential operators $A$, $B$, $C$ and $D$ are defined in Eqs. (\ref{3-40}) - (\ref{3-45}). These operators depend on the equation of state $P(\mu,T)$ and the normal fluid density $n_n(\mu,T)$. Both quantities have to be provided by an underlying microscopic theory. The structure of the eigenvalue problem is similar to the simple zero temperature case. We will show below that the zero temperature limit of Eq. (\ref{2-9}) is indeed given by Eq. (\ref{2-2}) and this behavior will also be found in the frequencies.

We see that calculating the frequencies at non-vanishing temperature is as straightforward as it was at $T=0$. However, results on the temperature dependence of hydrodynamic collective modes are rare in the literature \cite{Ho1}. This is because the equation of state and normal fluid density are in general not known or only in certain ranges of temperature and chemical potential. But when applying the local density approximation $\mu \rightarrow \mu - V_{ext}(r)$ one drives through a wide interval of values for the chemical potential and therefore a complete equation of state in $\mu$ has to be provided. Our motivation is in applying the Functional Renormalization Group, which is capable of providing the full thermodynamics of cold quantum gases, especially the functions $P(\mu,T)$ and $n_n(\mu,T)$. For the three-dimensional weakly interacting Bose gas the equation of state has been calculated by Lee and Yang \cite{LY1}. Their formulas are valid for small gas parameter $n a^3$ for temperatures not too close to the critical temperature. We comment on the critical behavior later in this section. The Lee--Yang equation of state is summarized in App. \ref{appLY}. We neglect the next-to-leading order LHY-correction (\ref{2-3}) because we are interested in thermal effects. 

In \cite{Ho1} Shenoy and Ho have calculated the temperature behavior of collective modes of a trapped Bose gas from two-fluid hydrodynamics and Lee--Yang theory for $0.6 < T/T_c < 1.2$. The authors considered the range of parameters where in addition to $na^3 \ll 1$ also $a n \lambda_T^2 \ll 1$ holds. ($\lambda_T = (2\pi\hbar^2/m k_B T)^{1/2}$ is the thermal wavelength.) In this limit the integral expressions for the equation of state from Lee--Yang theory can be evaluated analytically. We will, however, keep the full expressions for the equation of state.

\begin{figure}[tb!]
\includegraphics[scale=1,keepaspectratio=true]{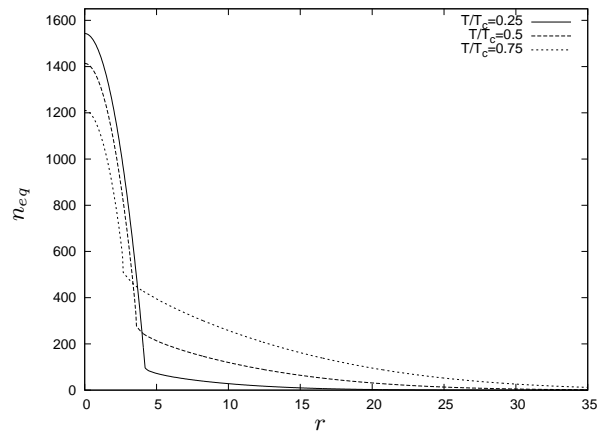}
\caption{Equilibrium density profile as a function of the radius obtained from Lee--Yang theory using local density approximation for a spherical symmetric potential. We observe a peak in the center of the trap. In the outer regions the gas is in its normal phase. }
\label{fig4}
\end{figure}

Before discussing the collective modes of a Bose gas at non-vanishing temperature we comment on some aspects of the static configuration of the trapped gas which are necessary for the interpretation of our results. For a homogeneous system with temperature $T$ we can calculate the critical chemical potential $\mu_c(T)$ of the superfluid phase transition. Within the local density approximation in the trapped system there will be a radius $r_c$ corresponding to the phase boundary between the superfluid and the normal regions of the cloud. This radius fulfills $\mu_c = \mu(0) - V_{ext}(r_c)$ with the chemical potential $\mu(0)$ in the center of the trap. The characteristic picture of the density profile consists in a narrow condensate peak in the center of the trap which is surrounded by a broad thermal cloud of the normal gas. Of course, there is also a non-vanishing contribution of the normal component to the inner regions. We visualize this behavior in Fig \ref{fig4}.

It is now apparent that for the description of a trapped gas at \emph{any} nonzero temperature one has to know the equation of state for both the superfluid and normal phase. In particular the presence and dynamics of the normal gas are of importance. There can be oscillations of the thermal cloud itself and, furthermore, it provides a non-trivial background for the oscillations of the condensate. As we approach zero temperature the thermal cloud vanishes. Condensate oscillations correspond to solutions $\delta n$ of the hydrodynamic equations which are only nonzero inside a sphere with radius $r_c$.

\begin{figure*}[tb!]
\includegraphics[scale=1,keepaspectratio=true]{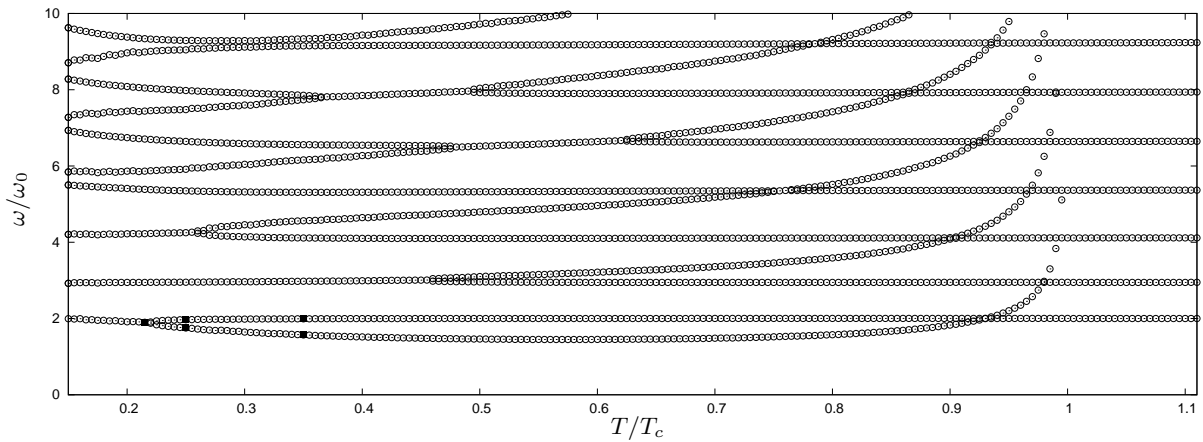}
\caption{Oscillations frequencies of collective modes ($l=0$) of both the condensate and thermal cloud in an isotropic harmonic trap. The equation of state is provided by Lee--Yang theory. In this plot the particle number is fixed to be $N=10^6$ for each value of $T/T_c$. The oscillations of chemical potential and temperature which correspond to the black solid dots are shown in Figs. \ref{fig9} and \ref{fig10}.}
\label{fig6}
\end{figure*}

The critical temperature $T_c$ of the trapped gas is defined as the temperature where the condensate peak appears in the center of the cloud. This can be reformulated as $\mu(0)=\mu_c$ or equivalently $r_c=0$. For Lee--Yang theory the critical temperature is given by
\begin{equation}
 \label{2-11} T_c = 2 \pi \left(\frac{\mu(0)}{2 g \zeta(3/2)}\right)^{2/3}.
\end{equation}
If we set $\mu(0)=\mu_c$ and calculate the number of particles $N$ in the trap, we find the ideal gas prediction $T_c^{(0)} (N)= \hbar \omega_0 (N/\zeta(3))^{1/3}$ to be satisfied for small gas parameter. There are several possibilities to change the parameter $T/T_c$. For fixed $\mu(0)$, the particle number will change when decreasing the temperature. Another way of going to low temperatures is to keep the particle number fixed. Then the chemical potential in the center of the trap has to be adjusted appropriately. If non-destructive measurements of frequencies become possible, a particular promising experimental way would be a change of $T/T_c$ by a smooth variation of the trap frequency, keeping the number of trapped atoms fixed.

In Fig. \ref{fig6} the isotropic ($l=0$) temperature dependent oscillation frequencies in an isotropic harmonic trap are shown for fixed particle number $N=10^6$ and $a/\ell_0=0.0005$, where $\ell_0=\sqrt{\hbar/m\omega_0}$ is the oscillator length. This particular choice of parameters implies a small gas parameter $na^3$, which is required in order to use Lee--Yang theory, but does not guarantee the applicability of hydrodynamics. Nevertheless, we expect this approach to capture the characteristics of the temperature dependence of hydrodynamic collective modes. Due to thermodynamic constraints the equation of state for the strongly coupled case will share common features with Lee--Yang theory and thus will show a qualitatively similar oscillation spectrum. For a quantitatively precise description of the collective modes of the weakly coupled system, which is not the purpose of this paper, one needs to include corrections to the hydrodynamic behavior by means of collisionless dynamics. Our results then constitute a limiting case.

We observe a rich spectrum of frequencies. The oscillations can be classified as condensate oscillations and oscillations of the thermal cloud. The condensate oscillations correspond to the branches which disappear at $T_c$. The computability of frequencies above the critical temperature by our method is a manifestation of the fact that the Landau two-fluid model remains formally valid for vanishing superfluid component. Of course, the frequencies for $T \geq T_c$ could be computed equivalently with a one-fluid model for the thermal liquid. We find the oscillation frequencies of the thermal cloud to be very close to the spectrum which one obtains using the equation of state for an ideal Bose gas and vanishing superfluid density. In particular, the lowest mode which also exists above $T_c$ is given by $\omega= 2 \omega_0$, see for example Ref. \cite{Str2}.

For low temperatures ($T/T_c \lesssim 0.1$) the two-fluid region, defined by $r \leq r_c$ with $r_c$ from above, starts to grow considerably and the normal fluid in the outer region vanishes. However, we were not able to obtain the eigenfrequencies of this crossover numerically in a sufficiently accurate manner. In order to get an understanding of the very low temperature limit of the collective modes we therefore used an implementation where the cloud is cut off at $r_c$ and thus only the two-fluid region is left. One then obtains a spectrum of the type shown in Fig. \ref{fig7}, which shows a smooth limit for $T \rightarrow 0$, as is discussed below. (For increasing $T$ there remains a systematic error in Fig. \ref{fig7} because with increasing temperature the normal mantle surrounding the two-fluid core of the cloud cannot be neglected anymore.)

From Fig. \ref{fig6} it is clear that there are level-crossing frequencies over wide ranges of $T/T_{c}$. These branches can be used as a thermometer by measuring a certain number of frequencies and then locating them in the plot. How this can be done experimentally by response techniques is explained in Sec. \ref{measure}.

\begin{figure}[tb!]
\includegraphics[scale=1,keepaspectratio=true]{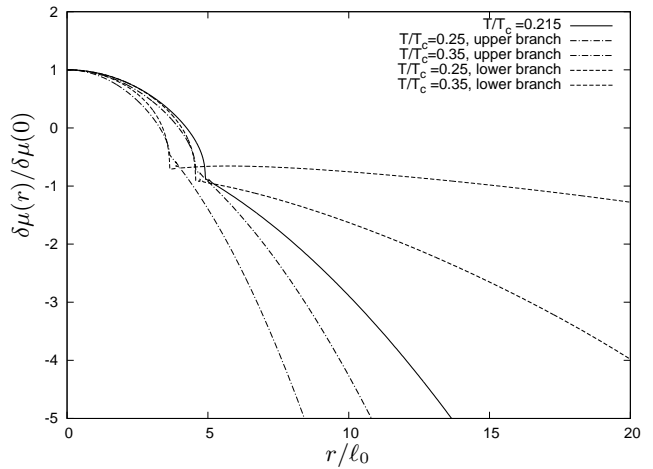}
\caption{Oscillation profile of the chemical potential for the frequencies corresponding to the the black solid dots in Fig. \ref{fig6}. For $T/T_c=0.215$ there is only one solution, which then splits up into a lower branch indicated by the dashed line, and an upper branch represented by the dot-dashed line. The thinner lines correspond to higher temperatures. We observe the lower branch to form a bump in the region of the condensate, while the upper branches smoothes out this boundary. The eigenfunctions are normalized here to be $1$ in the center of the trap.}
\label{fig9}
\end{figure}

\begin{figure}[tb!]
\includegraphics[scale=1,keepaspectratio=true]{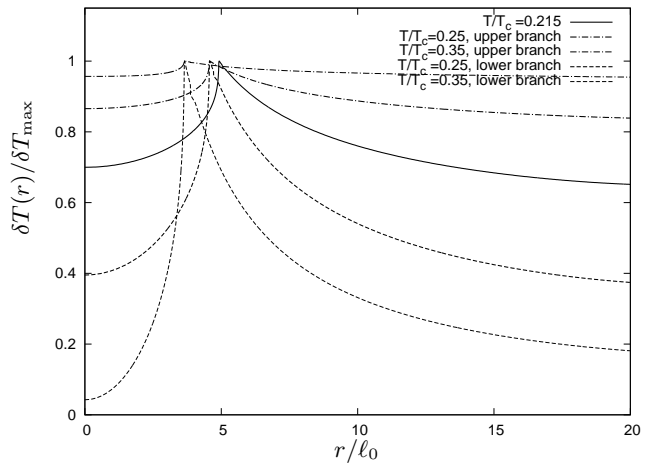}
\caption{The same setting as in Fig. \ref{fig9} but showing the oscillation profile of temperature. Starting from a solution which possesses a cusp, the lower branch pronounces this peak while the oscillations corresponding to the upper branch become flat. The eigenfunctions are normalized here such that they have value $1$ at the peak.}
\label{fig10}
\end{figure}

\begin{figure}[tb!]
\includegraphics[scale=1,keepaspectratio=true]{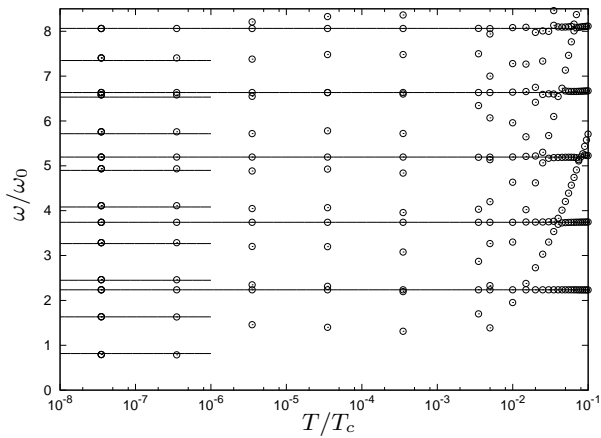}
\caption{Zero temperature limit of the collective oscillations. Here, the chemical potential is kept fixed at $\mu(0)/\hbar \omega_0=10$, which does not coincide with the choice of parameters in Fig. \ref{fig6}. The solid lines over the whole range of temperatures correspond to formula (\ref{2-2b}) for superfluid oscillations at $T=0$. The frequencies given by Eq. (\ref{2-15}) are indicated by solid lines only for small temperatures, where this formula is expected to be valid. In this figure, there is a systematic error for high temperatures, because the thermal cloud is not treated appropriately. }
\label{fig7}
\end{figure}

An apparent feature of Fig. \ref{fig6} is the appearance of bifurcations points where one branch of collective excitation frequencies splits up into two distinct branches. The meaning of these points is very intuitive. While for low temperatures superfluid and normal fluid component are closely connected, the modes split up at intermediate values of $T/T_c$. One of them represents the superfluid oscillation, which vanishes at the critical temperature, the other one corresponds to the oscillation of the normal fluid which remains present also above $T_c$. The black solid dots in Fig. \ref{fig6} correspond to points before and after a splitting. We indeed find that for low temperatures there is only one solution $\delta \mu$ and $\delta T$ for this particular eigenvalue, which then smoothly splits up into a set of two solutions with different properties. We make this statement clear in Figs. \ref{fig9} and \ref{fig10} where the (isotropic) oscillations of chemical potential and temperature are shown for the frequencies corresponding to the black dots in Fig \ref{fig6}, distinguished by the labels ``upper'' and ``lower'' branch, meaning higher and lower frequency, respectively. 

One has to keep in mind, that the oscillations of particle number density and entropy density, $\delta n$ and $\delta s$, vanish at infinity because the thermodynamic functions in Eqs. (\ref{3-30}) and (\ref{3-31}) vanish. However, the oscillations of density and entropy density are related to $\delta \mu$ and $\delta T$ via
\begin{equation}
\label{2-11b}\left(\begin{array}{c}
       \delta n \\ \delta s
      \end{array}
\right)
=
 \left(\begin{array}{cc}
	P_0^{\mu\mu} & P_0^{\mu T} \\ P_0^{\mu T} & P_0^{TT}
      \end{array}
\right)
\left(\begin{array}{c}
       \delta \mu \\ \delta T
      \end{array}
\right),
\end{equation}
where the matrix contains the higher derivatives of the equation of state $P(\mu,T)$ applied to local density approximation. (We use here the notation of Sec. \ref{coll}.) Since this matrix vanishes at infinity, $\delta \mu$ and $\delta T$ can be nonzero. At zero temperature, where we have a sharp cloud radius of the condensate, the fluctuations of the chemical potential $\delta \mu$ are discontinuously going to zero at this radius.

From our discussion of the equilibrium density profile we see that for all temperatures below the critical temperature $T_c$ there is a radius $r_c$ where the critical equation of state $P(\mu_c,T)$ has to be known. The vicinity of the point $\mu_c$ might be small but the behavior of the thermodynamic functions in this interval could nevertheless influence the frequencies significantly. Indeed, the superfluid phase transition is of second order and we expect the specific heat at constant pressure $C_P = T(\partial S/\partial T)_{P,N}$ and the isothermal compressibility $\kappa_T = -V^{-1} (\partial V/\partial P)_{T,N}$ to be singular at $\mu_c$. Their behavior in the critical region as a function of temperature is $C_P \sim |T-T_c|^{-\alpha}$ and $\kappa_T \sim |T-T_c|^{-\gamma}$ with the critical indices \cite{PV} of the three-dimensional XY-universality class. The specific heat shows a cusp while the isothermal compressibility diverges. A systematic study of these effects on the collective modes would be very interesting. Lee--Yang theory does not allow for such an investigation.

Using the Lee-Yang formulas given in App. \ref{appLY} we can calculate numerically the coefficient functions appearing in the operators $A$, $B$, $C$, and $D$ in Eq. (\ref{2-9}). For very low temperatures we observe that $B$ and $C$ vanish while $A$ approaches its zero temperature expression, given by Eq. (\ref{2-2a}),
\begin{equation}
 \label{2-12b} 
\left(\begin{array}{cc}
	A & B \\ C & D
      \end{array}
\right)
\stackrel{T \rightarrow 0}{\longrightarrow}
\left(\begin{array}{cc}
	A(T=0) & 0 \\ 0 & D(T=0)
      \end{array}
\right).
\end{equation}
This corresponds to a decoupling of superfluid and thermal degrees of freedom. This is expected to happen in the zero temperature limit since the few atoms in the thermal cloud have no influence on the dynamics of the condensate. The operator entering the eigenvalue problem is now block diagonal and all zero temperature frequencies are reproduced. They correspond to Stringari's mean field formula given above. (Note that we have neglect the LHY-correction.) It is interesting that $D$ does not vanish. Indeed, the coefficient functions $a_1$, $d_1$ and $d_2$ entering $A$ and $D$ from Eqs. (\ref{3-43}) and (\ref{3-45}) satisfy
\begin{equation}
 \label{2-13} \frac{a_1}{d_1} \stackrel{T\rightarrow 0}{\longrightarrow} \frac{1}{3}, \hspace{5mm} d_2 \stackrel{T\rightarrow 0}{\longrightarrow} -\frac{1}{6}.
\end{equation}
This can be seen by evaluating the integrals for $P(\mu,T)$ and $n_n(\mu,T)$ numerically or applying an expansion for $T \ll \mu$ using only the phonon part of the spectrum, see App. \ref{exact}. It can be shown that $a_1/d_1$ corresponds to the ratio of the squared first and second velocities of sound, $c_1^2$ and $c_2^2$, respectively. Therefore we find the well-known relation 
\begin{equation}
 \label{2-14} c_2^2 = \frac{c_1^2}{3} 
\end{equation}
to be satisfied for very low temperatures. 

We show in App. \ref{exact} that the behavior of the coefficient functions of $D$ as $T\rightarrow 0$, Eq. (\ref{2-13}), can be used to derive the eigenvalues of $D$ by slight modifications from the eigenvalues of $A$, which are given by Eq. (\ref{2-2b}) with $\alpha=1$. The corresponding eigenfrequencies are found to be
\begin{equation}
 \label{2-15} \omega_{nl} = \left(\frac{4n(n+l)-l}{6}\right)^{1/2} \omega_0,\mbox{ } (n \geq1).
\end{equation}
The experimental verification of this formula will be difficult, however. Although for low $T$ (and $T\rightarrow 0$) these frequencies are solutions to the two-fluid hydrodynamic equations, there are only very few atoms left in the normal phase to oscillate. From Eq. (\ref{2-15}) it is apparent that for $n=1, l=0$ one obtains $\sqrt{2/3}\omega_0$ which is \emph{below} the trapping frequency. This cannot be achieved by formula (\ref{2-2b}) with an effective polytropic index $\alpha_{eff}$. The existence of such modes below the trapping frequency might be a special feature of superfluid hydrodynamics just as the appearance of a second velocity of sound. Indeed, in their early experiments Stamper-Kurn et al. \cite{SK1} used a cigar-shaped harmonic potential with trapping frequency $\omega_{0z} / 2 \pi =  18.04(1)$ Hz in axial direction. They observed an out-of-phase oscillation between condensate and thermal cloud of a Bose gas at $T = 1 \mu K$ in the hydrodynamic regime. The critical temperature was around $1.7 \mu K$. The measured frequency of axial motion was found to be $\omega_0 / 2 \pi = 17.26(9)$ Hz, which is below $\omega_{0z}$.

In Fig. \ref{fig7} we show our numerical results for the frequencies in the zero temperature limit using the full Lee--Yang equation of state for $\mu(0)/\hbar\omega_0=10$ fixed and $a/\ell_0=0.0005$. We find formulas (\ref{2-2b}) and (\ref{2-15}) to be in perfect agreement with our data.

\subsection{Measuring the spectrum with response techniques}
\label{measure}
We have seen that information is contained both in the precise values of single modes and in the whole spectrum of the oscillating system. While the lowest lying modes can be excited by varying the parameters of the trapping potential (e.g. the potential depth) and then measuring the frequency in a free expansion, the whole spectrum could be obtained by an other method. In principle one can fit a superposition of arbitrary many modes to the freely expanding cloud but the contribution of the higher modes will only be weak. We suggest to use a response measurement instead to locate the frequencies of the trapped cloud. In order to keep the notation short we restrict this discussion to the zero temperature case.

The eigenfrequencies and corresponding eigenfunctions obtained so far represent normal coordinates of an oscillating system, the trapped cold gas. The equation of motion $\partial_t^2 \delta \mu + A \delta \mu=0$ has solutions $\delta \mu_{nlm}=\bar{g}_{nl}(r) r^l Y_{lm} e^{-i \omega_{nl} t}$ which are harmonic oscillations with frequency $\omega_{nl}$. (More precisely, one should use the operator $E$ defined below Eq. (\ref{e2}) instead of $A$. This is of no relevance here, because they have the same eigenvalues.) 

To describe an experimental situation we need to include dissipation effects into our description. We therefore introduce a phenomenological damping term by replacing $\partial^2_t \rightarrow \partial^2_t + \Gamma \partial_t$, with damping constant $\Gamma$. The new eigenfrequencies of the system will be denoted by $\Omega_{nl}$. We use Fourier decomposition and write $\delta \mu(t) = \int \mbox{d}\Omega \delta \mu(\Omega) e^{-i\Omega t}$. Since we already know the eigenvalues of $A$ we immediately obtain the quadratic equation
\begin{equation}
 \label{2-15c} (-\Omega^2_{nl} - i \Gamma \Omega_{nl} +\omega_{nl}^2) \delta \mu_{nlm}(\Omega) =0
\end{equation}
for possible solutions $\Omega_{nl}$. The latter simplify due to the fact that $t_2=1/\Gamma$ represents a characteristic time scale over which damping takes place. In order to observe oscillations this scale should be substantially larger than $t_1=1/\omega_{nl}$. The other choice would correspond to the overdamped case. Hence $\omega_{nl}^2 \gg \Gamma^2$ and we have
\begin{equation}
 \label{2-15d} \Omega_{nl} \simeq  \omega_{nl}\left(1-\frac{\Gamma^2}{8 \omega_{nl}^2}\right) - \frac{i \Gamma}{2}
\end{equation}
for the new eigenfrequencies of the system. Due to damping effects they acquire an imaginary part and their real part is slightly shifted. This phenomenological treatment gives no explanation of the microscopic nature of the damping terms. It would be interesting to derive the relation between the damping constants $\Gamma$ and the five transport coefficients $\eta, \kappa_T,\zeta_1, \zeta_2$ and $\zeta_3$ of two-fluid hydrodynamics \cite{Put1}. In principle, one expects $\Gamma$ to depend on the mode in the form of some continuous function $\Gamma(\Omega)$. For simplicity, we use here the same phenomenological $\Gamma$ for all modes.

The picture of the trapped gas as an oscillating system will provide us with an intuition how individual (and thus also higher lying) modes can be addressed by virtue of response techniques. Consider a classical point particle with Hamiltonian $H=\frac{1}{2m}p^2 + \frac{m}{2} \omega^2x^2$. With a damping term included its equation of motion will have the form of Eq. (\ref{2-15c}) and is therefore independent of the mass of the particle. However, when applying a small external force $f(t)$ the equation of motion for the Fourier components of $x(t)$ becomes
\begin{align}
 \label{2-15e} (-\Omega^2 - i \Gamma \Omega + \omega^2) x(\Omega) = \frac{1}{m} f(\Omega).
\end{align}
Obviously, the response $\chi(\Omega) = x(\Omega)/f(\Omega)$ is proportional to the inverse of the mass $m$. It is therefore more difficult to excite a heavy particle than a lighter one. We introduce an analog quantity $m_{nlm}$ which represents the ``mass'' for the collective oscillation $\delta \mu_{nlm}$. We call $m_{nlm}$ the ``response coefficients''. The relation between the amplitude of the oscillations of the atom cloud and the oscillation of the external perturbation $f(\Omega)$ is governed by $m_{nlm}$.

Let us consider a periodic perturbation of the harmonic trapping potential,
\begin{equation}
 \label{2-15f} V_{ext}(\vec{x}) \rightarrow V_{ext}(\vec{x}) + \mbox{cos}(\Omega t) F(r) r^l Y_{lm}.
\end{equation}
The perturbation is assumed to be small in comparison to $V_{ext}$. As we show in App. \ref{appE} this small variation acts as a driving force on the oscillating system (\ref{2-15c}). There it is also shown that in order to excite an oscillation with angular dependence $\delta \mu_{nlm} \propto Y_{lm}$, $\delta V_{ext}$ has to have the same angular dependence. In Eq. (\ref{mass}) we give a general formula for the response coefficient $m_{nlm}$ of the collective mode. Given $\Omega \simeq \omega_{nl}$ we then have
\begin{equation}
 \label{2-15h} \mbox{Im}\chi(\Omega) = \frac{1}{m_{nlm}} \frac{\Gamma \Omega}{(\Omega^2-\omega_{nl}^2)^2+\Gamma^2 \Omega^2}
\end{equation}
for the imaginary part of the response function defined in Eq. (\ref{chi}). The latter is related to dissipation of energy in the system. 

\begin{figure}[tb!]
\includegraphics[scale=1,keepaspectratio=true]{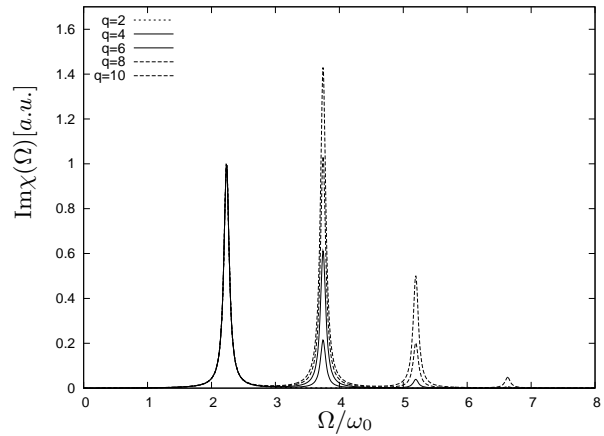}
\caption{Response of the zero temperature mean-field Bose gas to a perturbation $\delta V_{ext} =r^q \mbox{cos}(\Omega t) $ of the external potential. }
\label{fig8}
\end{figure}

For a demonstration we restrict ourselves to the case of a three-dimensional mean-field Bose gas, i.e. $P(\mu) \propto \mu^2$, and consider only the isotropic modes with $l=0$. In addition we assume a monomial perturbation $F(r) = r^q$ with $q \geq 0$. This includes the cases of a spatially constant background field $\delta V_{ext}(\vec{x},t) = \mbox{cos}(\Omega t)$ ($q=0$) and a time-variation of the trapping frequency, $\delta V_{ext} =r^2 \mbox{cos}(\Omega t) $ ($q=2$). Interestingly, we find in App. \ref{appE} that not all values of $q$ can be used to excite the whole spectrum. For example, there is no response at all to $q=0$, while for $q=2$ only the breathing mode will be excited and for $q=4$ only the lowest two modes oscillate. We show this behavior in Fig. \ref{fig8}. (In our intuitive picture developed above the absence of response corresponds to an infinite mass.) The normalization in Fig. \ref{fig8} is such that the lowest mode always has the same amplitude. The relative amplitudes, both for oscillations with different frequencies, and for oscillations with a given frequency, but for different excitations, can be used as a further observable for testing the system. The computation of ratios of amplitudes for different $\Omega$ needs knowledge of $\Gamma(\Omega)$. In principle, this can be measured from the width of the resonance. Ratios of amplitudes for fixed $\Omega$ and different radial or angular dependence of the perturbation are independent of $\Gamma(\Omega)$ and $m_{nlm}$. In principle, amplitudes as in Fig. \ref{fig8} can be measured both by dissipation ($\mbox{Im} \chi$) or by the amplitude of the cloud oscillations ($\mbox{Re}\chi$).

We conclude that the variation of the external potential with a certain frequency and radial and angular dependence allows for an excitation of individual modes. The oscillation can then be detected by imaging the density profile or, hopefully, by new non-invasive methods.

\subsection{Outlook on other systems}
\label{outlook}
We mentioned in the beginning of this section that we expect the features of the Bose gas to be generic for other systems of cold atoms. At nonzero temperature this requires the applicability of the two-fluid model which is related to a two-component order parameter. The latter can be written as $\phi(\vec{x}) = |\phi(\vec{x})| e^{i \theta(\vec{x})}$ at each space-time point. If $\theta(\vec{x})$ is varying slowly in space one can define the superfluid velocity as its gradient, $\vec{v}_s \propto \nabla \theta$. Thus our description remains valid for systems which are described by a complex scalar field which acquires a non-vanishing expectation value. This includes in particular composite bosons like the atom-atom dimers on the BEC-side of the BEC-BCS crossover. For an imbalanced system it would then be interesting to include the conserved spin degrees of freedom of the remaining fermions as an additional macroscopic variable in the hydrodynamic equations.

At zero temperature \emph{any} system with hydrodynamic equations of motion might be considered. An example can be found in Ref. \cite{Cit1} where the collective oscillation frequencies of a one-dimensional dipolar quantum gas have been calculated.

The polytropic equation of state $P \propto \mu^{\alpha+1}$ is of particular interest for scale-invariant systems like ultracold Fermi gases at the unitary point where the scattering length $a$ diverges. At zero temperature only the chemical potential remains as a dimensionful quantity, the pressure has to be proportional to $\mu^{(d+2)/2}$. Indeed, introducing the Bertsch parameter $\xi$ we can write the chemical potential at the unitary point as $\mu = \xi \varepsilon_F$, where $\varepsilon_F \propto n^{2/d}$ is the Fermi energy density. Using $\mbox{d}P = n \mbox{d} \mu$ we arrive at $P(\mu) \propto \mu^{(d+2)/2}$ as claimed. This time we know the equation of state exactly, but only for zero temperature. At the unitary point and for $T=0$ the oscillation frequencies obey the exact Eq. (\ref{2-2b}), with $\alpha=3/2$ for a three-dimensional spherical trap.

Around the unitary point and in the dilute, three-dimensional regime, the equation of state can be parametrized as \cite{Bul1}
\begin{equation}
\label{2-16} \varepsilon(n) = \frac{3}{5} \varepsilon_F(n) \left( \xi - \frac{\zeta}{k_F(n)a} + O\left(\frac{1}{(k_Fa)^2}\right)\right)
\end{equation}
with an additional parameter $\zeta$ implying a shift
\begin{equation}
\label{2-17} \frac{\delta \omega_{B}^2}{\omega_{B}^2} = \frac{256}{525 \pi} \frac{\zeta}{\xi} \frac{1}{k_F(n(0)) a}
\end{equation}
of the breathing mode.

\section{Collective modes - general theory}
\label{coll}

In this section we derive the eigenvalue problem (\ref{2-9}) for collective modes at $T\geq0$ from Landau's two-fluid ideal hydrodynamics \cite{Lan1,Put1}. The heart of every hydrodynamic description lies in the expansion of general constitutive equations in terms of derivatives of the contributing fields. The range of applicability of this description has already been discussed in the introduction. The constitutive relations consist in conservation laws for particle number, momentum, energy and additional conserved quantities. The lowest order of this expansion is static equilibrium and thus thermodynamics. The next order is called ideal hydrodynamics and describes non-equilibrium processes without dissipation. The latter is included in higher orders of the expansion.

\subsection{The eigenvalue problem from two-fluid hydrodynamics}
Observations on quantum fluids suggest that there is a critical temperature $T_{c}$ such that for $0 \leq T \leq T_{c}$ the macroscopic hydrodynamic motion can be divided into two kind of flows, a superfluid (subscript $s$) and a normal fluid (subscript $n$) one. Their velocity fields $\vec{v}_s(\vec{x},t)$ and $\vec{v}_n(\vec{x},t)$, respectively, enter the momentum density $\vec{g}=n_s \vec{v}_s + n_n \vec{v}_n$ with coefficients satisfying $n=n_s + n_n$, where $n(\vec{x},t)$ is the density of the gas or fluid under consideration.  We might be lead to the conclusion that the substance is built out of two different  kinds of fluids. However, the true nature of superfluidity consists in quantum physics and our description is only formally correct. The two flows can be distinguished by the following properties. The superfluid velocity field is a gradient field and hence irrotational. Furthermore, entropy is only carried by the normal fluid part. The full hydrodynamic equations under these constraints were set up by Lev. D. Landau in a famous and still very worth reading paper in 1941 \cite{Lan1}. We concentrate on ideal hydrodynamics where dissipation terms are neglected and entropy is conserved. In the homogeneous case, the underlying conservation laws read
\begin{align}
\label{3-1} \frac{\partial n}{\partial t} + \mbox{div}(\vec{g}) &=0,\\
\label{3-2} \frac{\partial g_i}{\partial t} + \partial_k \Pi_{ik} &=0,\\
\label{3-3} \frac{\partial s}{\partial t} + \mbox{div}(s \vec{v}_n) &=0,\\
\label{3-4} \frac{\partial \vec{v}_s}{\partial t} + \nabla\left(\frac{v_{s}^2}{2} + \mu\right) &=\vec{0}
\end{align}
with stress tensor $\Pi_{ik} = n_n v_{n,i} v_{n,k} + n_s v_{s,i} v_{s,k} +  P \delta_{ik}$, entropy density $s$, pressure $P = -\varepsilon + T s + \mu n + n_n (\vec{v}_n - \vec{v}_s)^2$, chemical potential $\mu$ and internal energy density in the rest frame of the gas, $\varepsilon$. These equations have to be supplemented by the equation of state $P(\mu,T)$ and the normal fluid density $n_n(\mu,T)$. For dissipative hydrodynamics one would replace Eq. (\ref{3-3}) by the conservation of energy and include an additional term in Eq. (\ref{3-4}). We use dimensionless units described in appendix \ref{unit}, setting $\hbar=k_B=m=1$.

In order to understand what changes when applying an external field, we first derive the \textit{local density approximation} from a thermodynamic point of view. For this purpose consider a gas contained in a potential $V_{ext}(\vec{x})$ such that local equilibrium is reached at each point of space. If we take two neighboring small but yet macroscopic parts of the gas their energies $E_1$ and $E_2$ and particle numbers $N_1$ and $N_2$ will adjust in a manner maximizing the entropy $S=S_1+S_2$ under the constraints $E_1+E_2=E$ and $N_1+N_2=N$, respectively. This implies temperature and the \textit{full} chemical potential to be constant inside the trap. However, the full chemical potential is given by the Gibbs free energy per particle and thus reads $\mu_{full}(\vec{x}) = \mu_{hom}(P(\vec{x}),T) + V_{ext}(\vec{x})$, where $\mu_{hom}(P,T)$ is the equilibrium chemical potential of the homogeneous system as a function of pressure $P$ and temperature $T$. We conclude that a system where the chemical potential $\mu$ in the homogeneous equilibrium functions is substituted according to $P_{hom}(\mu,T) \rightarrow P_{hom}(\mu-V_{ext},T)$ etc. behaves like a system trapped in an external potential of large spatial extend. This rule is called local density approximation. 

An external potential can be included in Eqs. (\ref{3-2}) and (\ref{3-4}) by virtue of force terms,
\begin{align}
\label{3-5} \frac{\partial g_i}{\partial t} + \partial_k \Pi_{ik} &= - n \partial_i V_{ext},\\
\label{3-6} \frac{\partial \vec{v}_s}{\partial t} + \nabla\left(\frac{v_{s}^2}{2} + \mu\right) &= - \nabla V_{ext}.
\end{align}
The static solution denoted by a subscript $0$ gives the expected expressions $\mu_0(\vec{x}) + V_{ext}(\vec{x}) = \bar{\mu}$ and $\nabla P_0(\vec{x}) + n_0(\vec{x}) \nabla V_{ext}(\vec{x}) = \vec{0}$ of static equilibrium. Using the Gibbs-Duhem relation $\mbox{d} P = s \mbox{d}T + n \mbox{d} \mu$ and thus $\nabla P_0 = s_0 \nabla T_0 + n_0 \nabla \mu_0$ at every space point, we conclude $s_0 \nabla T_0 =\vec{0}$ in equilibrium. We recognize the right thermodynamic behavior to emerge naturally. Pressure and densities of mass, entropy and energy are constant in time but space dependent, while temperature is constant all over the trap and the chemical potential follows the rule of local density approximation. The parameter $\bar{\mu}$ can be adjusted in order to get a certain  particle number $N$. The fluid velocities $\vec{v}_s$ and $\vec{v}_n$ vanish in equilibrium.

The next step towards our eigenvalue problem consists in expanding the two-fluid equations in small fluctuations around their static equilibrium solution such that time-dependent quantities only appear linearly. We write 
\begin{align}
\label{3-7} \mu(t,\vec{x}) &= \mu_0(\vec{x}) + \delta \mu(t,\vec{x}),\\
\label{3-8} T(t,\vec{x}) &= T + \delta T(t,\vec{x}),\\
\label{3-9} \vec{v}_{s,n}(t,\vec{x}) &= \delta \vec{v}_{s,n}(t,\vec{x}).
\end{align}
In the same way we linearize $P=P_0 + \delta P, n = n_0 + \delta n, s= s_0 +\delta s$. Eqs. (\ref{3-1}), (\ref{3-3}), (\ref{3-5}) and (\ref{3-6}) then become
\begin{eqnarray}
\label{3-10}\partial_t \delta n +\mbox{div}(\delta \vec{g}) &= 0,\\
\label{3-11}\partial_t \delta\vec{v}_s + \nabla \delta \mu &=\vec{0},\\
\label{3-12}\partial_t \delta \vec{g} + \nabla \delta P + \delta n \nabla V_{ext} &=\vec{0},\\
\label{3-13}\partial_t \delta s + \mbox{div}(s_0 \delta\vec{v}_n) &=0
\end{eqnarray}
with $\delta \vec{g}=n_{s,0} \delta\vec{v}_s + n_{n,0} \delta\vec{v}_n$. Using the relation $\nabla P_0 = s_0 \nabla T_0 + n_0 \nabla \mu_0$ we get $\nabla \delta P = s_0 \nabla \delta T + \delta n \nabla \mu_0 + n_0 \nabla \delta \mu = s_0 \nabla \delta T - \delta n \nabla V_{ext} + n_0 \nabla \delta \mu$, since $\nabla T_0 =\vec{0}$ and $\nabla \mu_0 = - \nabla V_{ext}$. This can be used to cast Eq. (\ref{3-12}) into the form
\begin{equation}
\label{3-14} n_{n,0} \partial_t \left(\delta \vec{v}_n - \delta \vec{v}_s\right) + s_0 \nabla \delta T =\vec{0}.
\end{equation}

In the following we will consider a spherically symmetric harmonic trapping potential $V_{ext}(\vec{r})=\frac{m}{2} \omega_0^2 r^2$. The generalization of our formalism to other radial potentials $V_{ext}(\vec{r}) = V_{ext}(r)$ is straightforward. In Eq. (\ref{3-34}) we give a formulation of the eigenvalue problem which holds for arbitrary external potentials $V_{\rm ext}(\vec{x})$ and thus also allows for a treatment of anisotropic trapping geometries.

We assume the fluctuations of the physical quantities to be a harmonic oscillation in time with frequency $\omega$,
\begin{equation}
\label{3-15} \delta \mu(\vec{x},t) = e^{- i \omega t}\delta \mu(\vec{x}),
\end{equation}
and similar expressions for the other time-dependent functions. The remaining spatial deviations are again denoted by $\delta \mu$ etc. Since time has completely dropped out of our system of partial differential equations, this slight abuse of notation hopefully does not lead to confusion. We recognize our obtained set of equations
\begin{align}
\label{3-16} i \omega \delta n &= \mbox{div}(n_{s,0}\delta \vec{v}_s + n_{n,0} \delta \vec{v}_n),\\
\label{3-17} i \omega \delta \vec{v}_s &= \nabla \delta \mu,\\
\label{3-18} i \omega (\delta \vec{v}_n - \delta \vec{v}_s) &= \frac{s_0}{n_{n,0}} \nabla \delta T,\\
\label{3-19} i \omega \delta s &= \mbox{div}(s_0 \delta \vec{v}_n)
\end{align}
to be an eigenvalue problem for a differential operator acting on $\delta \mu$, $\delta T$, $\delta \vec{v}_s$ and $\delta \vec{v}_n-\delta\vec{v}_s$.

\subsection{Zero temperature}
We derive the zero temperature eigenvalue problem in terms of an equation of state given in the form $P(\mu)$. We introduce the notation $\mu_0(r)=\mu(0)-V_{ext}(r)=\mu(0)-\omega_0^2 r^2/2$. Since we have set $\hbar=k_B=m=1$, we can express all dimensionful quantities with respect to $\omega_0$, see App. \ref{unit}. In the following we set $\omega_0=1$ in all formulas, which means that we are measuring time in units of the inverse trapping frequency. Thus, $\mu_0(r)=\mu(0)-r^2/2$. Furthermore, we define
\begin{equation}
\label{3-20} P^{\mu}_0(r) = \left(\frac{\partial P}{\partial \mu}\right) (\mu_0(r)) \mbox{, } P^{\mu \mu}_0(r) = \left(\frac{\partial^2 P}{\partial \mu^2}\right) (\mu_0(r)),
\end{equation}
where $\mu(0)$ is the chemical potential in the center of the trap. We can write $\delta n(\vec{x}) = P^{\mu \mu}_0(r) \delta \mu(\vec{x})$ because $\mbox{d}P = n \mbox{d}\mu$. There are no temperature fluctuations present at $T=0$ and thus the only degrees of freedom are $\delta \mu$ and $\delta \vec{v}_s$ described by Eqs. (\ref{3-16}) and (\ref{3-17}). The latter equations can be decoupled. Writing $n_0(r)=P^{\mu}_0(r)$ we arrive at the Stringari wave equation \cite{Str1}
\begin{equation}
\label{3-21} \omega^2 P_0^{\mu\mu} \delta\mu + \mbox{div}(P_0^\mu \nabla \delta \mu)=0.
\end{equation}
So far, $\delta \mu$ depends on the $d$-dimensional spatial coordinate $\vec{x}$. Since we are dealing with a spherically symmetric trap we can classify the possible solutions to the eigenvalue problem via an ansatz
\begin{equation}
\label{3-22} \delta \mu(\vec{x}) = \bar{g}(r) r^l f_{lm},
\end{equation}
where $f_{lm}$ are spherical harmonics. (For a definition of spherical harmonics in $d\leq 3$ dimensions we refer to App. \ref{appC}. Note that we have $l=0,1$ for $d=1$.) When applying this ansatz to Eq. (\ref{3-21}) we benefit from the facts that $\nabla \mu_0 \cdot \nabla f_{lm}=0$ and $\partial_r P(\mu_0(r))=-rP_0^\mu(r)$. In addition we use the relations
\begin{align}
\label{3-23} &\vec{e}_r \cdot \nabla \delta \mu(\vec{x}) = \left(\bar{g}'(r)+ \frac{l}{r} \bar{g}(r)\right) r^l f_{lm},\\
\label{3-24} &\Delta \delta \mu(\vec{x}) = \left(\bar{g}''(r) + \frac{2l + d -1}{r}\bar{g}'(r)\right)r^lf_{lm},
\end{align}
where $\vec{e}_r$ denotes the unit vector pointing in radial direction. We arrive at
\begin{eqnarray}
\nonumber 0 &=& \omega^2 \bar{g}(r) - r \bar{g}'(r) - l \bar{g}(r)\\ 
\label{3-25} &&+ \frac{P^\mu_0(r)}{P^{\mu\mu}_0(r)} \left(\bar{g}''(r) + \frac{2 l + d -1}{r}\bar{g}'(r)\right).
\end{eqnarray}
When substituting $z=r^2$ the eigenvalue problem is given by
\begin{equation}
 \label{3-26} A g(z) = \omega^2 g(z)
\end{equation}
for $g(z) =\bar{g}(r)$ and the differential operator
\begin{equation}
 \label{3-27} A = -\frac{P^\mu(z)}{P^{\mu\mu}(z)} \left(4 z \frac{\partial^2}{\partial z^2} + 2(2 l +d)\frac{\partial}{\partial z}\right) + \left(2 z \frac{\partial}{\partial z} +l \right).
\end{equation}
This equation has to be fulfilled on the interval $z\in[0,R^2]$ where $R$ is the radius of the static cloud defined by $n_0(r=R)=0$. We do not allow for fluctuations to change the cloud radius since this would be a second order small effect. Since we have set $\omega_0$ to $1$, only $\omega^2$ appears in Eq. (\ref{3-26}) instead of $\omega^2/\omega_0^2$.

\subsection{Non-vanishing temperature}
For temperatures $T \geq 0$ we have to implement temperature fluctuations and the full thermodynamic functions $P(\mu,T)$ and $n_n(\mu,T)$. Extending the notation introduced in Eq. (\ref{3-21}) we write
\begin{equation}
\label{3-26b} P^{\mu T}_0(r) = \frac{\partial^2 P}{\partial T \partial \mu}(\mu_0(r),T).
\end{equation}
In the same manner $P^\mu_0$, $P^{\mu\mu}_0, P^T_0$, $P^{TT}_0$, $\tilde{n}_0$, and $\tilde{n}_0^\mu$ are defined for $T\geq0$, where 
\begin{equation}
\label{3-27b} \tilde{n} = \frac{s^2}{n_n}.
\end{equation}
The static solutions $n_0$ and $s_0$ are connected to $\mu_0(r)=\mu(0)-r^2/2$ and $T_0=T$ via $n_0=P^\mu_0, s_0=P^T_0$. Working again with independent variables $\delta \mu$ and $\delta T$ we get
\begin{align}
\label{3-28} \delta n = P^{\mu\mu}_0 \delta \mu + P^{\mu T}_0 \delta T,\\
\label{3-29} \delta s = P^{T\mu}_0 \delta \mu + P^{T T}_0 \delta T.
\end{align}
From our analysis at zero temperature we know that due to the spherically symmetry the eigenmodes are most easily obtained by eliminating the velocity fields. We find
\begin{align}
\label{3-30} \omega^2 \delta n + \mbox{div}(n_0 \nabla \delta \mu + s_0 \nabla \delta T) &=0,\\
\label{3-31} \omega^2 \delta s + \mbox{div}(s_0 \nabla \delta \mu + \tilde{n}_0 \nabla \delta T) &=0,
\end{align}
or in terms of $P(\mu,T)$,
\begin{align}
\label{3-32}\omega^2 P^{\mu\mu}_0 \delta \mu + \omega^2 P^{\mu T}_0 \delta T +\mbox{div}(P^\mu_0\nabla \delta \mu + P^T_0 \nabla \delta T ) &=0,\\
\label{3-33}\omega^2 P^{T \mu}_0 \delta \mu + \omega^2 P^{T T}_0 \delta T +\mbox{div}(P^T_0 \nabla \delta \mu + \tilde{n}_0 \nabla \delta T ) &=0
\end{align}
determining $\omega$. Note that $\tilde{n}_0$ supplements the equation of state in the latter equation. We have chosen $\tilde{n}_0$ instead of $n_{n0}$ in order to keep the notation simple. 

Eqs. (\ref{3-32}) and (\ref{3-33}) can be written in the canonical form of an eigenvalue problem. Indeed, they are equivalent to
\begin{align}
\nonumber
\left(\begin{array}{cc}
	P^{\mu\mu}_0 & P^{\mu T}_0 \\ P^{\mu T}_0 & P^{TT}_0
      \end{array}
\right)^{-1}
\left(\begin{array}{cc}
	-\mbox{div}(P^\mu_0\nabla \cdot) & -\mbox{div}(P^T_0\nabla \cdot) \\ -\mbox{div}(P^T_0\nabla \cdot) & -\mbox{div}(\tilde{n}_0\nabla \cdot)
      \end{array}
\right)
\left(\begin{array}{c}
       \delta \mu \\ \delta T
      \end{array}
\right)\\
\label{3-34} = \omega^2
\left(\begin{array}{c}
       \delta \mu \\ \delta T
      \end{array}
\right),
\end{align}
where we assumed $P^{\mu T}_0 = P^{T \mu}_0$. The first matrix appearing in this equation is the inverse of the Hessian matrix of $P(\mu,T)$. The latter is positive for thermodynamic reasons and thus the inverse always exists. The external trapping potential $V_{\rm ext}(\vec{x})$ enters Eq. (\ref{3-34}) through the application of the local density approximation indicated by the subscript $0$.

Analogously to the zero temperature case we classify the solutions for an isotropic harmonic potential $V_{\rm ext}(\vec{x}) = r^2/2$ via an ansatz
\begin{align}
\label{3-37}\delta \mu (\vec{x}) &= g(r^2) r^l f_{lm},\\
\label{3-38}\delta T(\vec{x}) &= h(r^2) r^l f_{lm}
\end{align}
where $l=0,1,2,...$ and $f_{lm}$ are the corresponding spherical harmonics. The resulting equations simplify on going over to the new variable $z=r^2$. We arrive at 
\begin{equation}
\label{3-39}
\left(\begin{array}{cc}
	A & B \\ C & D
      \end{array}
\right)
\left(\begin{array}{c}
       g(z) \\ h(z)
      \end{array}
\right)
= \omega^2
\left(\begin{array}{c}
       g(z) \\ h(z)
      \end{array}
\right),
\end{equation}
where the operators $A, B, C, D$ are defined by
\begin{align}
\label{3-40} A = &a_1(z) \left(4 z \frac{\partial^2}{\partial z^2} + 2(2 l +d)\frac{\partial}{\partial z}\right) + \left(2 z \frac{\partial}{\partial z} +l \right)\\
\label{3-41} B = &b_1(z) \left(4 z \frac{\partial^2}{\partial z^2} + 2(2 l +d)\frac{\partial}{\partial z}\right) + b_2(z) \left(2 z \frac{\partial}{\partial z} +l \right)\\
\label{3-42} C = &c_1(z) \left(4 z \frac{\partial^2}{\partial z^2} + 2(2 l +d)\frac{\partial}{\partial z}\right)\\
\label{17} D = &d_1(z) \left(4 z \frac{\partial^2}{\partial z^2} + 2(2 l +d)\frac{\partial}{\partial z}\right) + d_2(z) \left(2 z \frac{\partial}{\partial z} +l \right)
\end{align}
with $z$-dependent coefficients
\begin{align}
\label{3-43}  &a_1 =\frac{P^{\mu T}_0 P^T_0 - P^{TT}_0 P^{\mu}_0}{\mbox{det}P_0}, 
&b_1 =\frac{P^{\mu T}_0 \tilde{n}_0 - P^{TT}_0 P^{T}_0}{\mbox{det}P_0},\\
\label{3-44}  &b_2 =\frac{P^{TT}_0 P^{\mu T}_0 - P^{\mu T}_0 \tilde{n}^\mu_0}{\mbox{det}P_0},
&c_1 =\frac{P^{\mu T}_0 P^{\mu}_0 - P^{\mu \mu}_0 P^T_0}{\mbox{det}P_0},\\
\label{3-45}  &d_1 =\frac{P^{\mu T}_0 P^T_0 - P^{\mu \mu}_0 \tilde{n}_0}{\mbox{det}P_0},
&d_2 =\frac{P^{\mu \mu}_0 \tilde{n}^\mu_0 - (P^{\mu T}_0)^2}{\mbox{det}P_0}
\end{align}
with $\mbox{det}P_0=P^{\mu \mu}_0 P^{TT}_0-(P^{\mu T}_0)^2$. Again, $\omega=\sqrt{l}$ is a solution for constant $g(z)$ and vanishing $h(z)$, which is independent of the equation of state. 

An additional complication to the case of vanishing temperature arises from the fact that the normal part of the background density $n_{n0}$ is not restricted to a region $z \leq z_{max}$ as it is the case for the superfluid density $n_{s0}$. In contrast it decreases exponentially for large $z$ (c.f. Fig. \ref{fig4}). In praxis, we expect that the region $z>R^2$ for a sufficiently large value of $R^2$ is not affecting the oscillation frequencies since only an exponentially small number of particles is in that region. We restrict our numerical treatment therefore to a finite sphere $z\leq R^2$.

Given $a_1,...,d_2$ over the whole range $z \in [0,R^2]$ the problem of determining the collective frequencies is as straightforward as it was for $T=0$. Indeed, as we show in Sec. \ref{num} a simple and yet efficient method to solve for the eigenvalues will be given by replacing the operators by matrices. Thus, if the more complicated coefficient functions (\ref{3-43}) - (\ref{3-45}) are given to a sufficient degree of accuracy, the finite temperature collective modes are easily obtained.

\subsection{Numerical implementation}
\label{num}
Now that we have identified the ordinary differential operator(s) describing two-fluid hydrodynamic modes, we present a numerical method to determine the corresponding eigenfrequencies. The input for our approach consists in the equation of state $P(\mu,T)$ and normal fluid density $n_n(\mu,T)$ on the interval $\mu\in[\mu_{min},\mu(0)]$. Since we restricted ourselves to the case of spherically symmetric trapping the partial differential hydrodynamic equations in $d$ variables $x_1,\dots,x_d$ are projected onto an ordinary differential equation in $z$. This is a disadvantage of our method when compared to the fact that most experiments are performed in asymmetric traps. 

In order to introduce our numerical procedure we consider the case of $T=0$ since the operators for $T\geq0$ have an identical structure but different coefficient functions. We solve the eigenvalue problem (\ref{2-2}), (\ref{2-2a}) by discretizing the interval $z\in[0,z_{max}]$ via $i=0,1,\dots,M$ and $\Delta z = z_{max}/M$ such that $z_i=i \Delta z$. The function $g(z)$ becomes a vector $g = (g(z_i))_{i=0,\dots,M}$ and $A$ is represented by a matrix $(A_{ij})$ having components
\begin{align}
\nonumber &A_{ij} = 2 z_i \frac{\delta_{i+1,j}-\delta_{i-1,j}}{2 \Delta z} + l \delta_{ij}-\frac{P^\mu_0(z_i)}{P^{\mu\mu}_0(z_i)}\\
\label{3-57} &\times \left(4 z_i \frac{\delta_{i+1,j}+\delta_{i-1,j}-2 \delta_{ij}}{\Delta z^2}+ 2(2 l +d) \frac{\delta_{i+1,j}-\delta_{i-1,j}}{2 \Delta z}\right)
\end{align}
for $i=1,...,M-1$ and $j=0,...,M$ with $\delta_{ij}$ being the Kronecker delta. The cases $i=0$ and $i=M$ require some care. We define
\begin{align}
\label{3-58} A_{0,j} = &l \delta_{0,j} -\frac{P^\mu_0(0)}{P^{\mu\mu}_0(0)} 2(2 l+ d) \frac{\delta_{1,j}-\delta_{0,j}}{\Delta z},\\
\nonumber A_{M,j} = &2z_{max} \frac{\delta_{M,j}-\delta_{M-1,j}}{\Delta z} + l \delta_{M,j}\\
\label{3-59} &-\frac{P^\mu_0(z_{max})}{P^{\mu\mu}_0(z_{max})} 2(2 l +d) \frac{\delta_{M,j}-\delta_{M-1,j}}{\Delta z}
\end{align}
having replaced the second-order difference scheme by a first order one and set the second derivative $g''$ at the boundary $z=z_{max}$ to zero. The eigenvalues of $(A_{ij})$ can now be determined using standard procedures. 

Our discretization of $A$ is closely related to the underlying physical problem. The functions $g(z)$ on which $A$ operates describe macroscopic hydrodynamic fluctuations of the chemical potential. Thus they are implicitly assumed to be finite and free of a rich local structure including jumps and kinks. This applies in particular to the lowest lying modes with only a few radial nodes. Therefore we can safely assume $g(z)$ to be sufficiently smooth such that the approximations $g'(z_i) = (g_{i+1}-g_{i-1})/(2\Delta z) + O(\Delta z^2)$ and $g''(z_i) = (g_{i+1}-2g_i+g_{i-1})/(\Delta z^2) + O(\Delta z^2)$ hold in the open interval $(0,z_{max})$. However, at the points $z_0=0$ and $z_M=z_{max}$ we run into problems as $z_{-1}$ and $z_{M+1}$ are not defined. A first guess might be to assign certain boundary values to $g$. But investigating the breathing mode  for $\alpha=1$ and $d=3$ in Eq. (\ref{b6}), $g(z)=1-5z/6\mu(0)$ for $z\leq z_{max}$, $0$ otherwise, we recognize none of the sensible boundary conditions $g(z_{max})=0$ or $g'(z_{max})=0$ to be satisfied. This does not come as a surprise since our description necessarily breaks down in the outer regions of the cloud where the static density goes to zero. The crucial point is to prevent the solution from diverging at the boundaries. So we demand it to be Taylor expandable even there, although the physical solution being zero for $z>z_{max}$ might have a discontinuity, and replace the second-order difference scheme in $z_{-1}$ and $z_{M+1}$ by a first-order one, i.e. for $z_0=0$
\begin{equation}
\label{3-60} \frac{g_{1} - g_{-1}}{2 \Delta z} = \frac{g_1 - g_0}{\Delta z}+O(\Delta z).
\end{equation}
This can be solved for $g_0$ and yields $g_0 \simeq (g_1+g_{-1})/2$. The latter mean-value property is true for any function which can be linearized around $z=0$ provided a sufficiently small step size. In the same way, the approximation of $g'(z_{max})$ in Eq. (\ref{3-59}) can be justified. Setting $g''(z_{max})=0$ in the same equation implies $g_{M+1}-2g_M+g_{M-1}=0$ in a discretized version, which leads us to the same mean-value property.

This simple approach is very efficient. As an example we show in Tab. \ref{tabPol} the first ten eigenfrequencies corresponding to a polytropic equation of state $P(\mu) \propto \mu^{\alpha+1}$ for different choices of $\alpha$, $d$ and $l$ compared to the exact solution given by Eq. (\ref{2-2b}). The underlying grid of points is of size $M=2000$. We recognize the (more interesting) lower lying frequencies to converge faster. Of course, the accuracy can be improved by further increasing $M$.
\begin{table}
\begin{tabular}{|c|c||c|c||c|c|}
\hline
(i) exact & (i) num. & (ii) exact & (ii) num. & (iii) exact & (iii) num.\\
\hline
3.46410 & 3.46410 & 1.50214 & 1.50214 & 1.00000 & 1.00000\\
8.00000 & 7.99982 & 2.35339 & 2.35337 & 2.16025 & 2.16025\\
12.4900 & 12.4894 & 3.13786 & 3.13777 & 3.10913 & 3.10913\\
16.6706 & 16.9693 & 3.89609 & 3.89591 & 4.00000 & 4.00000\\
21.4476 & 21.4452 & 4.64095 & 4.64064 & 4.86484 & 4.86484\\
25.9230 & 25.9191 & 5.37802 & 5.37752 & 5.71548 & 5.71548\\
30.3974 & 30.3916 & 6.11010 & 6.10936 & 6.55744 & 6.55744\\
34.8712 & 34.8631 & 6.83880 & 6.83775 & 7.39369 & 7.39369\\
39.3446 & 39.3335 & 7.56510 & 7.56367 & 8.22598 & 8.22598\\
43.8178 & 43.8031 & 8.28963 & 8.28773 & 9.05539 & 9.05539\\
\hline
\end{tabular}
\caption{Exact and numerical frequencies in units of $\omega_0$ obtained for an equation of state $P=\mu^{\alpha+1}$ for (i)  $\alpha=0.1$, $d=1$, $l=0$, (ii) $\alpha=3.9$, $d=1$, $l=0$ and (iii) $\alpha =3$, $d=3$, $l=1$. All results correspond to a grid of $M=2000$ points.}
\label{tabPol}
\end{table}

For a particular physical situation it may be favorable to use a \emph{non-uniform grid}, i. e. the points $z_i$ shall not be equidistant. Of course, the discretized operator $(A_{ij})$ can also be derived for a non-uniform grid. Such a discretization has been applied to calculate the data points for Fig. \ref{fig6}. The use of a finite grid can be accompanied by a systematic error which overestimates or underestimates the eigenfrequencies. Nevertheless, we believe our method to be reliable because the exact results in Eqs. (\ref{2-2b}) and (\ref{2-15}) are obtained to a sufficient accuracy and we found a quite fast convergence of the lowest eigenvalues from smaller to larger matrices.

For a Bose gas at nonzero temperature the outer regions of the cloud are in the normal phase and the density decays exponentially for large radii, $n \sim \mbox{Li}_{3/2}(e^{\beta \mu})$, where $\mbox{Li}_{\nu}(z)=\sum_k z^k/k^\nu$ is the polylogarithm. Our numerical solution on a finite grid requires an effective radius of the cloud although this is not a physical quantity. However, for too high distances from the center of the trap the gas will be so dilute that local equilibration cannot be ensured and therefore hydrodynamics does not provide the correct equations of motion for this part of the cloud. We therefore define an effective radius or minimal chemical potential such that $n(\mu_{\rm min}) \ell_0^3 =1$, where $\ell_0$ is the oscillator length. At this value of the chemical potential the interparticle spacing will be of the order of the oscillator length and this constitutes the crossover region where hydrodynamics breaks down. Since only very few atoms are in the outer regions of the cloud we expect the experimental hydrodynamic modes to be close to the modes which are calculated with this effective radius.

\section{Conclusions and outlook}
\label{concl}
In this paper we suggest that collective oscillations may ultimately be used as precision observables for trapped ultracold atom gases. Within the hydrodynamic approximation the frequencies of these oscillations directly reflect the thermodynamic equation of state. Their measurement can therefore shed light on the validity of non-perturbative methods for interacting many-body systems which are used for the computation of the equation of state. In parallel, further experimental advances may allow for a simultaneous measurement of several frequencies, which could permit the determination of thermodynamic variables, in particular the temperature, with high precision.

In the two-fluid hydrodynamic regime, the excitation and measurement of collective oscillations results in a set of discrete numbers which can be used to compare different theoretical methods. An experimental improvement of the precision of the measurement of collective modes would set stringent bounds on predictions from many-body calculations. Indeed, we have shown that the equation of state allows us to determine the oscillation frequencies as eigenvalues of a differential operator. Thus, if broad collections of data on collective oscillation frequencies would exist, this would necessarily rule out certain equations of state and theoretical approaches. In a certain sense this may be compared to restrictions of the composition and equation of state of the sun by helioseismology.

The dependence of the collective modes on system parameters as the scattering length is of particular interest. We have seen that interaction effects can lead to shifts of the zero temperature mean field results for the oscillation frequencies of a dilute Bose gas. For the three-dimensional case these shifts are rather small, however. Nevertheless, even there non-perturbative effects can be made visible. The appearance of an effective UV-cutoff related to the s-wave scattering approximation manifests itself in a dependence of the equation of state on the gas parameter and this cutoff. Beyond lowest order perturbation theory we describe the effects of the LHY-correction for the oscillation spectrum. We also explore the range of larger scattering length, where perturbation theory is no longer valid, by means of an equation of state from Functional Renormalization Group (FRG). For gas parameters $n(0) a^3$ above $10^{-4}$ the difference between various methods for the relative frequency shift reaches the percent level and may be measurable.

The analysis of the dilute Bose gas in two dimensions revealed further interesting many-body effects. Using again the equation of state from FRG there are substantial deviations of the collective modes from mean field theory for large values of the dimensionless coupling constant. In these cases the difference of the breathing mode from its mean field value was found to be of order $1 \% - 10 \%$ and showed an extremum for a specific value of the chemical potential in the center of the trap. Very interesting would be the investigation of these effects in a dimensional crossover from three to two dimensions.

A main result of this paper concerns the temperature dependence of the oscillation frequencies. We suggest that this could ultimately be used as a precision thermometer. As an example we have computed the temperature dependent oscillation spectrum based on an equation of state obtained from Lee--Yang theory. The collective oscillations of the condensate and the thermal cloud show a rich spectrum with non-trivial dependence on the temperature, including level-crossing and the disappearance of the superfluid modes at the critical temperature. 

The treatment of the oscillation spectrum at non-vanishing temperature has to account for the coexistence of two interacting fluids, one for the superfluid condensate and the other for the thermal cloud of uncondensed atoms. The thermal cloud disappears for $T \rightarrow 0$, while the condensate vanishes as the critical temperature is approached, $T \rightarrow T_c$. Any computation of the spectrum has to reproduce these limiting cases correctly. In particular, we show how the zero temperature limit of the two-fluid hydrodynamic equations has a solution corresponding to a zero temperature formula for the condensate oscillations given by Stringari. We also derived a new analytic result for frequencies at very low temperatures which are carried by the components of the thermal cloud. Our method of calculation covers the critical temperature as well as $T > T_c$, where the thermal cloud is the only fluid component. It will be interesting to study the effects of critical physics related to the second order superfluid phase transition on the oscillation frequencies.

In order to measure not only the lowest collective oscillation frequencies but rather parts of the whole spectrum we suggest to use response techniques and excite frequencies individually. Once not only the lowest lying frequencies are accessible within an appropriate accuracy, but also higher modes, these will allow for precise verifications of the consistency of theoretical models.

The formal developments of this paper concern a method for the computation of oscillation frequencies in a trap, given the equation of state and normal fluid density for a homogeneous liquid. Starting from Landau's ideal two-fluid hydrodynamics we have derived the general expression for a differential operator whose input is the equation of state given as pressure $P(\mu,T)$ and normal fluid density $n_n(\mu,T)$. The eigenvalues of this operator are the collective oscillation frequencies of the trapped gas. For simplicity we restricted ourselves to the case of an isotropic, $d$-dimensional, harmonic trapping. While the extension to an arbitrary spherically symmetric trapping potential $V_{ext}(r)$ is straightforward, anisotropy would invalidate our symmetry assumptions. Experiments are often performed in not completely spherically symmetric arrangements. We expect our findings to be qualitatively correct even for these cases. We propose a method for numerical implementation which is based on discretization. This can in principle also be done for an anisotropic trapping potential but with a higher computational effort.

The restriction to the case of ideal hydrodynamics can be released. To first order, our equations would be modified due to the five transport coefficients of superfluid hydrodynamics, including shear viscosity and heat conductivity. Here, immediately many interesting questions arise: How are these transport coefficients related to the damping rates of the modes, which appear as the phenomenological introduced widths in our treatment of linear response? Can we use collective modes for measurements on the transport coefficients, e.g. $\eta/s$? Is it possible to derive, say, $\eta(\mu,T)$ from a microscopic quantum field theory just like the equation of state $P(\mu,T)$?

We conclude that collective modes have a high potential to be used as precision observables, which can quantify different aspects of many-body physics. It seems worth to improve the techniques used for a precise measurement of oscillation frequencies. In particular, a non-invasive simultaneous measurement of several frequencies may allow for a precision estimate of the thermodynamic variables temperature and density or chemical potential.

\begin{appendix}
\section{System of units}
\label{unit}
In a relativistic system one usually sets $\hbar = k_B = c =1$ in order to express all quantities in terms of energies. However, for a non-relativistic system the velocity of light $c$ does not appear in the equations and it is preferable to set $\hbar = k_B = m =1$, where $m$ is a mass. (In our case this is the mass of the gas atoms.) All quantities can now be expressed in terms of a length scale of choice. The spherically symmetric trapping potential
\begin{equation}
\label{a1} V_{ext}(\vec{x}) = \frac{m}{2} \omega_0^2 r^2
\end{equation}
with $r^2 = x_1^2 + \dots + x_d^2$ provides such a scale given by the oscillator length $\ell_0=\sqrt{\hbar/m\omega_0}$. If all dimensionful quantities are expressed in terms of $\ell_0$, the latter will no longer appear in the equations. In other words, we can set $\ell_0=1$ in our formulas and understand all quantities to be measured in units of $\ell_0$. Since $\hbar=m=1$ this is equivalent to expressing everything in powers of $\omega_0$. Thus, the eigenvalues of $A g(z) = \omega^2 g(z)$ have to be understood as dimensionless quantities $\omega/\omega_0$.

We remark that due to $m=1$ mass density equals particle number density, $\rho =n$, which simplifies our hydrodynamic equations.

\section{Lee--Yang theory for Bose gas}
\label{appLY}
The equation of state for a weakly interacting Bose gas in three dimensions has been calculated by Lee and Yang in Refs. \cite{LY1}. Starting from the microscopic dispersion relation $\omega_p$ they compute the canonical partition function and derive the free energy density and other thermodynamic observables. The system is found to have two distinct phases, a normal and a superfluid one. Besides this Lee and Yang also deduce the two-fluid hydrodynamics for the weakly interacting Bose gas. These equations include that the mass $m^*$ of the collective excitations of the normal fluid component does not have to coincide with the mass $m$ of the atoms. For the present work we neglect such details and only give a short summary of the equation of state which enters the Landau two-fluid hydrodynamics.

Let density $n$ and temperature $T$ be given. We then have to determine the normal fluid density $n_n(n,T)$ such that
\begin{equation}
 \label{f1} n_n = \int_p \frac{1}{e^{\beta(\omega_p - \tilde{\mu})}-1},
\end{equation}
where $\int_p=\int \mbox{d}^3p/(2\pi)^{3/2}$, $\varepsilon_p=p^2/2$ and 
\begin{equation}
 \label{f2} \omega_p=\sqrt{\varepsilon_p(\varepsilon_p + 2g(n-n_n))}
\end{equation}
is the microscopic dispersion. Since the latter depends on $n_n$ itself, Eq. (\ref{f1}) has to be solved self-consistently. For zero temperature, where the normal fluid fraction vanishes, we arrive at the standard Bogoliubov dispersion relation $(\varepsilon_p(\varepsilon_p+2gn))^{1/2}$. We introduced a second function $\tilde{\mu}(n,T)$ and thus we have to provide an additional equation. For a critical density
\begin{equation}
 \label{f3} n_{c}(T)= \zeta(3/2)\left(\frac{T}{2 \pi}\right)^{3/2}
\end{equation}
this equation reads $n_n = n$ in the case of $n \leq n_{c}$, which corresponds to the normal phase. For $n\geq n_{c}$ we are in the superfluid phase and have
\begin{equation}
 \label{f4} \tilde{\mu} = g \int_p \frac{1}{e^{\beta(\omega_p - \tilde{\mu})}-1} \left(\frac{\varepsilon_p}{\omega_p}-1\right).
\end{equation}
The (more complicated) latter case with two equations for $n_n$ and $\tilde{\mu}$ can be solved numerically by iteration.

Once $n_n(n,T)$ and $\tilde{\mu}(n,T)$ are known the free energy density is found to be given by
\begin{equation}
 \label{f5} f(n,T) = \frac{g}{2}(n^2+n_n^2) + \tilde{\mu} n_n + T\int_p \mbox{log}(1-e^{-\beta(\omega_p-\tilde{\mu})}).
\end{equation}
From $\mu=(\partial f/\partial n)_T$ we derive the chemical potential
\begin{equation}
 \label{f6} \mu = \tilde{\mu}+ g(n+n_n).
\end{equation}
The pressure is then given by $P(\mu,T)=n\mu - f$, where $n(\mu,T)$ is found from inversion of Eq. (\ref{f6}). Notice that the simple (exact) relation between $n$ and $\mu$ for fixed $n_n$ and $\tilde{\mu}$ allows us to use the grand variables $\mu$ and $T$ right from the beginning. Then $n_n(\mu,T)$ and $\tilde{\mu}(\mu,T)$ have to be found self-consistently from Eqs. (\ref{f4}) - (\ref{f6}) using the dispersion relation
\begin{align}
\label{f2b} \omega_p&=
\left\{\begin{array}{cc}
	\sqrt{\varepsilon_p(\varepsilon_p + 2(\mu-\tilde{\mu}) -4g n_n)}, & \mu > \mu_{c} \\ \varepsilon_p, & \mu \leq \mu_{c}
      \end{array}
\right.
\end{align}
instead of Eq. (\ref{f2}). The critical chemical potential is given by $\mu_{c}=2 g n_{c}$.

The derivatives of the pressure can now be obtained in different ways. For the normal phase where the equations are particularly simple and involve only polylogarithmic functions one may use the canonical expressions derived from the free energy density and then insert $n(\mu,T)$ from Eq. (\ref{f6}). For example from the pressure $P(n,T)$ we can built $(\partial P/\partial n)_T = P^\mu/P^{\mu\mu}$. Since $P^\mu=n$ is known, we get an expression for $P^{\mu\mu}(\mu,T)$. Many expressions which are necessary for this process are given explicitly in the paper of Lee and Yang. For the superfluid  phase the authors also give formulas for certain ranges of temperature. However, we cannot rely on these limiting cases since we want to cover the full equation of state for all values of $\mu$ and $T$. We found that the numerical implementation of the integral expression of $P(\mu,T)$ derived from Eqs. (\ref{f5}) and (\ref{f6}) is suitable to calculate all other thermodynamic observables via finite difference formulas to sufficient accuracy.

\section{Spherical harmonics in one and two dimensions}
\label{appC}
We define spherical harmonics in $d$ dimensions for $l=0,1,2,...$ as complex valued functions $f_{lm}$ on the unit sphere $\mathbb{S}^{d-1}=\{\vec{x} \in \mathbb{R}^d | r^2=|\vec{x}|^2 = 1\}$ through the relation $\Delta (r^l f_{l m}) = 0$, where $\Delta$ is the Laplacian, $\Delta = (\partial/\partial x_1)^2+ \dots + (\partial/\partial x_d)^2$, and demand the orthonormality condition
\begin{equation}
\label{c1} \int_{\mathbb{S}^{d-1}} f^*_{l m} f_{l' m'} = \delta_{l l'} \delta_{m m'}.
\end{equation}
$m$ is a set of further indices. For $d=3$ we have $m \in \{-l,...,l\}$ and $f_{l m} = Y_{lm}$ in traditional notation. In $d=2$ dimensions, writing $x_1 = r \cos{\phi}, x_2=r \sin{\phi}$, we need to solve
\begin{equation}
\label{c2} 0 = \Delta(r^l f_{l m}(\phi)) = \left( \frac{\partial^2}{\partial r^2} + \frac{1}{r} \frac{\partial}{\partial r} + \frac{1}{r^2} \frac{\partial^2}{\partial \phi^2}\right) r^l f_{l m}(\phi),
\end{equation}
which leads to
\begin{equation}
\label{c3} f_{l, \pm 1} = \frac{1}{\sqrt{2 \pi}} e^{\pm i l \phi}
\end{equation}
for $l=0,1,2,..$. We see that all values of $l$ are allowed for $d=2$. This is no longer true in the one-dimensional case as can already be guessed by physical reasoning. For $d=1$ there are only monopole and dipole modes. Indeed, $\mathbb{S}^0=\{-1,1\}$ is a discrete point set and one-dimensional spherical coordinates read $x=\sigma r$ with $r=|x|$ and $\sigma=x/r$. The angular variable thus tells us whether we are considering the positive or the negative half line of the real axis. The expression $g(x)=g(r) f_l(\sigma)$ has to be understood as
\begin{equation}
\label{c4} g(x)= \left\{\begin{array}{cc}
	g(r) f_l(-1), & x<0 \\ g(0), & x=0 \\ g(r) f_l(1), & x>0.
      \end{array}
\right.
\end{equation}
Since the harmonic functions in $d=1$, i.e. the ones satisfying $g''(x) =0$, are exactly the affine linear functions, the relation $\Delta( r^l f_l(\sigma))=0$ can only be true for $l=0,1$. The normalization condition
\begin{align}
\nonumber \int_{\mathbb{S}^{0}} f^*_{l m} f_{l' m'} &= f^*_{l m}(-1) f_{l' m'}(-1) + f^*_{l m}(1) f_{l' m'}(1)\\
\label{c5} &= \delta_{l l'} \delta_{m m'}
\end{align}
is valid for
\begin{align}
\nonumber f_0(\sigma)&=
\left\{\begin{array}{cc}
	1/\sqrt{2}, & \sigma=-1 \\ 1/\sqrt{2}, & \sigma=1
      \end{array}
\right.,\\
\label{c6} f_1(\sigma)&=
\left\{\begin{array}{cc}
	-1/\sqrt{2}, & \sigma=-1 \\ 1/\sqrt{2}, & \sigma=1
      \end{array}
\right..
\end{align}
$f_0$ and $f_1$ defined in this way have the correct parity $(-1)^l$. We do not give explicit constructions of spherical harmonics for higher dimensions than three.

\section{Exact results at zero temperature}
\label{exact}
In this appendix we derive Eqs. (\ref{2-2b}) and (\ref{2-15}) for collective modes in the zero temperature limit. These calculations provide an intuition on how solutions to our proposed eigenvalue problem should look like. Moreover, the exact eigenfrequencies allow to perform perturbation theory beyond mean field. In such a way Eq. (\ref{2-5}) has been obtained in Ref. \cite{PiSt2}.

We start from the zero temperature eigenvalue problem $Ag=\omega^2 g$ with $A$ from Eq. (\ref{2-2a}). For a polytropic equation of state, $P(\mu) \propto \mu^{\alpha+1}$, the ratio $P_0^{\mu}(z)/P_0^{\mu\mu}(z)$ defined in Eq. (\ref{3-20}) is found to be
\begin{equation}
 \label{b1} \frac{P_0^\mu(z)}{P_0^{\mu\mu}(z)} = \frac{(\alpha+1)(\mu(0)-z/2)^{\alpha}}{(\alpha+1)\alpha(\mu(0)-z/2)^{\alpha-1}} = \frac{\mu(0)-z/2}{\alpha},
\end{equation}
where $\mu(0)$ is the chemical potential in the center of the trap. The eigenvalue problem then becomes
\begin{align}
 \nonumber 0 = &- \omega^2 g(z) + 2 z g'(z) + l g(z) \\
 \label{b2} &- \frac{\mu(0)-z/2}{\alpha}(4 z g''(z) + 2 (2l+d)g'(z)).
\end{align}
This equation has to be true for all values of $z$ inside the cloud. The radius of the cloud $R$ is found by solving $0=n_0(r=R)=(\alpha+1)(\mu(0)-R^2/2)^\alpha$, i.e. $z\leq R^2=2 \mu(0)$. Substituting $z \mapsto \bar{z}=z/2\mu(0)$, the chemical potential drops out,
\begin{align}
 \nonumber 0 = &-\omega^2 g(\bar{z}) + 2 \bar{z} g'(\bar{z}) + l g(\bar{z})\\
 \label{b3}  &- \frac{1-\bar{z}}{2 \alpha}(4 \bar{z} g''(\bar{z}) + 2 (2l+d)g'(\bar{z})).
\end{align}
As is pointed out in Ref. \cite{PeSm1} this is of the form of the hypergeometric differential equation. However, for arbitrary values of $\alpha$, $d$ and $l$ we would have to distinguish many special cases of its solution because the prefactors of the derivative terms in Eq. (\ref{b3}) change. Therefore, we apply a more direct method and ask for solutions of the form $g(\bar{z}) =\sum_{k\geq0} a_k \bar{z}^k$. This yields the recursion relation
\begin{equation}
\label{b4} a_{k+1}=\frac{\alpha ( l-\omega^2) + 2k(\alpha + k + l - 1 +d/2)}{2(k+l+d/2)(k+1)}a_k
\end{equation}
for the coefficients $a_k$. Since $A g(z)=\omega^2g(z)$ is invariant under a rescaling of $g(z)$ we can set $a_0=1$. A sufficient condition for the convergence of this series is termination and thus reduction to a polynomial of degree $n=0,1,...$ This will be the case for $k\leq n$ and
\begin{equation}
\label{b5} \omega_{\alpha,n,l} = \left( \frac{2n}{\alpha} (\alpha + n +l +d/2 - 1) + l\right)^{1/2} \omega_0,
\end{equation}
where we restored the unit $\omega_0$. Inserting this into Eq. (\ref{b4}) we get coefficients
\begin{equation}
\label{b6} a_{k+1}=\frac{(k-n)(\alpha+k+n+l+d/2 -1)}{(k+l+d/2)(k+1)}a_k
\end{equation}
and $a_0=1$.

In the zero temperature limit the eigenvalue problem (\ref{3-39}) at nonzero temperature gets block diagonal because $B$ and $C$ vanish. While $A$ approaches its zero temperature limit (\ref{2-2a}), the coefficient functions $d_1$ and $d_2$ of $D$ approach
\begin{align}
 \label{b7} &d_1 \stackrel{T\rightarrow 0}{\longrightarrow} - \frac{\tilde{n}_0}{P^{TT}_0},\\
 \label{b8} &d_2  \stackrel{T\rightarrow 0}{\longrightarrow} \frac{\tilde{n}^\mu_0}{P^{TT}_0}
\end{align}
with $\tilde{n_0}=(P^T_0)^2/n_{n,0}$, see Eq. (\ref{3-27b}). For the dilute Bose gas in three dimensions these ratios can be computed analytically from the formulas given in App. \ref{appLY}. In the zero temperature limit nearly all values of the chemical potential obtained via $\mu(z) = \mu(0) - V_{ext}(z)$ satisfy the condition $T \ll \mu(z)$. This will of course not be true for $\mu=0$ but this value can be neglected because it corresponds to the outermost points of the cloud. For $T \ll \mu$ the integrals for $P$ and $n_n$ receive contributions only from the phonon part of the spectrum. We have
\begin{align}
 \label{b9} P^T(\mu,T) &\simeq 4 T^3 \pi^2/(90 \mu^{3/2}),\\
 \label{b10} n_n(\mu,T) &\simeq 2 \pi^2 T^4/(45 \mu^{5/2}),
\end{align}
see for example \cite{PiSt1}. These quantities are powers of $T$ and will be extremely small. Nevertheless, the ratios (\ref{b7}) and (\ref{b8}) are non-vanishing because the powers of $T$ cancel each other. We find
\begin{align}
 \label{b11} \frac{\tilde{n}_0}{P^{TT}_0} &= \frac{\mu_0}{3} = \frac{\mu(0)-z/2}{3},\\
 \label{b12} \frac{\tilde{n}^\mu_0}{P^{TT}_0} &= -\frac{1}{6}
\end{align}
and the differential equation $D(T=0)h(z) = \omega^2 h(z)$ for thermal fluctuations becomes
\begin{align}
 \nonumber D(T=0)h(z)=&-\frac{1}{6} (2 z h'(z) +l h(z)) \\
 \nonumber &- \frac{\mu(0)-z/2}{3}(4 z h''(z) + 2(2l+3)h'(z))\\
 \label{b13}  =& \mbox{ }\omega^2 h(z).
\end{align}
We already inserted $d=3$. When we multiply both sides with $-6$ we find the same structure as in Eq. (\ref{b2}) but with the substitutions $\omega^2 \mapsto -6 \omega^2$ and $\alpha \mapsto -1/2$. Applying these modifications to the square of both sides of Eq. (\ref{b5}),
\begin{equation}
 \label{b14} \omega_{\alpha,n,l}^2 = \left(\frac{2n}{\alpha} \left(\alpha+n+l+\frac{1}{2}\right)+l\right) \omega_0^2,
\end{equation}
we find the eigenvalues of Eq. (\ref{b13}) to be
\begin{equation}
 \label{b15} \omega_{n,l} = \left(\frac{4n(n+l)-l}{6}\right)^{1/2} \omega_0.
\end{equation}
This proves Eq. (\ref{2-15}). The eigenfunctions read $h(z)=\sum_{k=0}^n a_k \bar{z}^k$ with $\bar{z}=z/2\mu(0)$, $a_0=1$ and
\begin{equation}
\label{b16} a_{k+1} = \frac{k(k+l) - n(n+l)}{(k+l+3/2)(k+1)}a_k.
\end{equation}

\section{Response to an external driving force}
\label{appE}
In addition to the trapping potential $V_{ext}(\vec{x}) = \frac{m}{2}\omega_0^2r^2$ we want to apply a small perturbation $\delta V_{ext}(\vec{x},t)$ such that the oscillating system experiences a driving force. We will show in this appendix how the equations of a driven oscillator arise and which perturbations can be used to excite the eigenfrequencies and thus measure the spectrum of collective modes. In order to keep the derivation short we only consider the zero temperature case. It is straightforward to generalize the following treatment to non-vanishing temperatures.

When we include the perturbing potential the linearized equation (\ref{3-11}) for the superfluid velocity becomes $\partial_t \delta \vec{v}_s + \nabla(\delta \mu + \delta V_{ext})=0$. Hence the Stringari wave equation (\ref{3-21}) is now replaced by
\begin{equation}
 \label{e1} (\partial_t^2 + \Gamma \partial_t) P^{\mu\mu}_0 \delta \mu - \mbox{div}(P^\mu_0\nabla \delta \mu) = \mbox{div}(P^\mu_0 \nabla \delta V_{ext}),
\end{equation}
where we also include damping like in Eq. (\ref{2-15c}). The left-hand side of this equation represents the unperturbed oscillator and the right-hand side corresponds to the external driving force. Applying a Fourier transformation $\delta \mu(\vec{x},t) = \int\frac{\mbox{d}\Omega}{2 \pi} e^{-i\Omega t} \delta\mu(\vec{x},\Omega)$ and analogous for $\delta V_{ext}(\vec{x},t)$ we obtain
\begin{equation}
 \label{e2} (-\Omega^2-i \Gamma \Omega + E) \delta \mu(\vec{x},\Omega) = - E \delta V_{ext}(\vec{x},\Omega),
\end{equation}
where we defined the operator $E = - (P_0^{\mu\mu})^{-1}\mbox{div}(P_0^\mu\nabla \cdot)$. Note that for $\phi_{nlm}(\vec{x})= g_{nl}(z)r^l f_{lm}$ we have $E \phi_{nlm}(\vec{x}) = r^l f_{lm} A g_{nl}(z) = \omega_{nl}^2 \phi_{nlm}(\vec{x})$ with $A$ from Eq. (\ref{3-27}) and corresponding eigenfunctions $g_{nl}(z)$ of $A$. Thus, $A$ and $E$ share the same eigenvalues. 

We introduce the scalar product $\left\langle f,g\right\rangle = \int_{\vec{x}} f(\vec{x})^* P^{\mu\mu}_0(r) g(\vec{x})$ where the integral extends over the cloud. The operator $E$ is then self-adjoint and its eigenfunctions $\phi_{nlm}$ form a complete and orthogonal system. For $f=g=\delta \mu$ we get $\left\langle f,g \right\rangle = \int \delta n \delta \mu$, where the chemical potential is multiplied by its thermodynamic conjugate, the density. We assume $||\phi_{nlm}||$ to be normalized to $1$ in the following.

For a perturbing potential $\delta V_{ext}(\vec{x},t)$, the response function $\chi$ is defined as
\begin{equation}
 \label{e3} \delta \mu(\vec{x},t) = \int_0^{\infty}\mbox{d}\tau \int_{\vec{y}} \chi(\vec{x},\vec{y},\tau) \delta V_{ext}(\vec{y},t-\tau).
\end{equation}
Introducing $\chi(\vec{x},\vec{y},\Omega)=\int_0^{\infty} \mbox{d}t e^{i\Omega t}\chi(\vec{x},\vec{y},t)$ this can be written as
\begin{equation}
 \label{e4} \delta \mu(\vec{x},\Omega) = \int_{\vec{y}} \chi(\vec{x},\vec{y},\Omega) \delta V_{ext}(\vec{y},\Omega).
\end{equation}
The spatial dependence on the coordinate $\vec{x}$ can be eliminated by expanding in the eigenfunctions $\phi_{nlm}$ of $E$, which yields
\begin{equation}
 \label{e5} \delta \mu_{nlm}(\Omega) = \sum_{n',l'm'} \chi_{nlm,n'l'm'}(\Omega) (\delta V_{ext})_{n'l'm'}(\Omega).
\end{equation}
From the equation of motion (\ref{e2}) we can immediately read off the response function in this particular basis. It is given by
\begin{equation}
 \label{e6} \chi_{nlm,n'l'm'}(\Omega) = \frac{-\omega_{nl}^2}{-\Omega^2-i \Gamma \Omega +\omega_{nl}^2} \delta_{nn'}\delta_{ll'}\delta_{mm'}.
\end{equation}
Eq. (\ref{e6}) shows that the mode $\phi_{nlm} e^{-i \omega_{nl} t}$ can be excited by applying a periodic perturbation of exactly this spatial form. 

For practical purposes we are more interested in the response of the trapped gas to an arbitrary perturbation such that in general several modes will be excited. Let the shape of $\delta V_{ext}$ be given by
\begin{equation}
 \label{e6b} \delta V_{ext}(\vec{x},t) = \frac{1}{2} \left( H(\vec{x}) e^{-i \Omega t} + H^*(\vec{x})e^{i \Omega t}\right).
\end{equation}
The variation of the external potential leads to a perturbation $\int_{\vec{x}} \delta n(\vec{x}) \delta V_{ext}(\vec{x},t)$ of the Hamiltonian of the system. The rate of average energy dissipated is therefore given by
\begin{equation}
 \label{e7} \frac{\mbox{d}E}{\mbox{d}t} = \left\langle \int_{\vec{x}} \delta n(\vec{x}) \partial_t \delta V_{ext}(\vec{x},t) \right\rangle,
\end{equation}
where the averaging is over a small interval of time. Using $\delta n = P^{\mu\mu}_0 \delta \mu$ this becomes
\begin{equation}
 \label{e7b} \frac{\mbox{d}E}{\mbox{d}t} = -\frac{\Omega}{2} \mbox{Im}\left\{ \int_{\vec{x},\vec{y}} H^*(\vec{x}) P^{\mu\mu}_0(r) \chi(\vec{x},\vec{y},\Omega) H(\vec{y})\right\}.
\end{equation}

For an oscillating system without spatial coordinate $\vec{x}$, Eq. (\ref{e7b}) would yield $\dot{E} = (\Omega/2) |H_0|^2\mbox{Im}  \chi(\Omega)$ for constant $H(\vec{x}) \equiv H_0$. If we consider the trapped system as a whole, we can use this analogy to define the response of the cloud to the perturbation $H(\vec{x})$. It has a clear physical meaning related to the dissipated energy. Expanding $H(\vec{x}) = \sum_{nlm} H_{nlm} \phi_{nlm}(\vec{x})$ we arrive at
\begin{equation}
 \label{e7c} \frac{\mbox{d}E}{\mbox{d}t} = \frac{\Omega}{2} \mbox{Im} \sum_{nlm} |H_{nlm}|^2 \frac{\omega_{nl}^2}{-\Omega^2-i \Gamma \Omega +\omega_{nl}^2}.
\end{equation}
Therefore, we have
\begin{equation}
 \label{chi} \chi(\Omega) = \sum_{nlm} \frac{1}{m_{nlm}} \frac{1}{-\Omega^2-i \Gamma \Omega +\omega_{nl}^2}
\end{equation}
with response coefficients
\begin{equation}
 \label{mass} \frac{1}{m_{nlm}} = |H_{nlm}|^2 \omega_{nl}^2.
\end{equation}

We simplify the situation even further by considering a three-dimensional mean-field Bose gas. As we have seen in Eq. (\ref{2-1}) this corresponds to an equation of state $P(\mu) = \mu^2/8\pi a$. The eigenfunctions were found in App. \ref{exact} to be given by $\phi_{nlm}(\vec{x}) = g_{nl}(\bar{z}) r^l f_{lm}$ with  $g_{nl}(\bar{z}) = \sum_{k=0}^n a_k \bar{z}^k$, $\bar{z}=(r/R)^2$ and $a_k$ defined in Eq. (\ref{b6}). The squared radius in this case is $R^2=2 \mu(0)$ with the chemical potential in the center of the trap $\mu(0)$. The lowest isotropic modes ($l=0$) are given by
\begin{align}
 \nonumber g_{10}(\bar{z}) &= 1 - \frac{5}{3} \bar{z},\\
 \nonumber g_{20}(\bar{z}) &= 1 - \frac{14}{3} \bar{z} + \frac{21}{5} \bar{z}^2,\\
 \nonumber g_{30}(\bar{z}) &= 1- 9 \bar{z} + \frac{99}{5} \bar{z}^2 - \frac{429}{35} \bar{z}^3,\\
 \label{e8} g_{40}(\bar{z}) &= 1 - \frac{44}{3} \bar{z} + \frac{286}{5} \bar{z}^2 - \frac{572}{7} \bar{z}^3 +\frac{2431}{63} \bar{z}^4.
\end{align}

We further assume $H$ entering Eq. (\ref{e6b}) to be of the form
\begin{equation}
 \label{e8b} H(\vec{x}) = F(r) r^L f_{LM}
\end{equation}
with monomial $F(r) = r^{q}$ and $q \geq 0$. We have $(m_{nlm})^{-1} \propto \delta_{lL}\delta_{mM}$. We conclude that a perturbing potential of the form (\ref{e8b}) has to have the right angular behavior ($\propto f_{lm}$) in order to excite a mode $\delta \mu \propto f_{lm}$. This is not restricted to the case of the Bose gas but holds in general. For $L=M=0$ we find
\begin{align}
 \nonumber \frac{1}{m_{100}(q)} &= \left(\frac{21 q \sqrt{5}}{2(q+3)(q+5)}\right)^2,\\
 \nonumber \frac{1}{m_{200}(q)} &= \left(\frac{165 (q-2)q\sqrt{14}}{8(q+3)(q+5)(q+7)}\right)^2,\\
 \nonumber \frac{1}{m_{300}(q)} &= \left(\frac{525 (q-4)(q-2)q\sqrt{27}}{16(q+3)(q+5)(q+7)(q+9)}\right)^2,\\
 \label{e9} \frac{1}{m_{400}(q)} &= \left(\frac{5985(q-6)(q-4)(q-2)q\sqrt{44}}{128(q+3)(q+5)(q+7)(q+9)(q+11)}\right)^2.
\end{align}
Here, we omitted an overall prefactor of $R^{3/2}/8\pi a$. The response coefficients depend strongly on $q$, i.e. the monomial character of $\delta V_{ext}$. In particular for low values of $q$ there will not be a response from the higher lying modes. For $q=0$, i.e. $\delta V_{ext}(t,\vec{x}) = \mbox{cos}(\Omega t)$, none of the collective modes will be excited, for $\delta V_{ext} =r^2 \mbox{cos}(\Omega t)$ only the breathing mode, and so on. We considered here isotropic oscillations. An analog behavior can be found for $l > 0$.

\end{appendix}

\end{document}